\documentclass[fleqn,a4paper,useAMS,usenatbib]{mnras}
\pdfoutput=1 
\usepackage{graphicx}	
\usepackage{amsmath}	
\usepackage{amssymb}	
\usepackage{multicol}        
\usepackage{bm}		
\usepackage{pdflscape}	
\usepackage{csvsimple}

\usepackage[T1]{fontenc}
\usepackage{ae,aecompl}

\usepackage{newtxtext,newtxmath}

\title[UV SEDs of early-type galaxies]{UV SEDs of early-type cluster galaxies: A new look at the UV upturn}

\author[S.~S. Ali et al.]{S.~S. Ali$^1$\thanks{Email: s.ali@bristol.ac.uk}, M.~N. Bremer$^1$, S.Phillipps$^1$, R. De Propris$^2$ \\
$^1$H.H.~Wills Physics Laboratory, University of Bristol, Tyndall Avenue, Bristol, BS8 1TL, UK. \\
$^2$FINCA, University of Turku, Turku, Finland. \\}

\begin{document}

\maketitle

\begin{abstract}
Using GALEX, UVOT and optical photometry, we explore the prevalence and strength of the UV-upturn in the spectra of quiescent early type galaxies in several nearby clusters. Even for galaxies with completely passive optical colours, there is a large spread in vacuum UV colour consistent with almost all having some UV upturn component. Combining GALEX and UVOT data below 3000\AA, we generate for the first time comparatively detailed UV SEDs for Coma cluster galaxies. Fitting the UV upturn component with a blackbody, twenty six of these show a range of characteristic temperatures (10000--21000K) for the UV upturn population. Assuming a single temperature to explain GALEX-optical colours could underestimate the fraction of galaxies with UV upturns and mis-classify some as systems with residual star formation. The UV upturn phenomenon is not an exclusive feature found only in giant galaxies; we identify galaxies with similar (or even bluer) $FUV-V$ colours to the giants with upturns over a range of fainter luminosities. The temperature and strength of the UV upturn are correlated with galaxy mass. Under the plausible hypothesis that the sources of the UV upturn are blue horizontal branch stars, the most likely mechanism for this is the presence of a substantial (between 4\% and 20\%) Helium rich ($Y > 0.3$) population of stars in these galaxies, potentially formed at $z\sim 4$ and certainly at $z>2$; this plausibly sets a lower limit of $\sim {\rm 0.3 - 0.8} \times 10^{10}$ $M_{\odot}$ to the {\it in situ} stellar mass of $\sim L^*$ galaxies at this redshift.\end{abstract}


\section{Introduction}
Despite their generally old and metal rich stellar populations, massive early-type galaxies often show excess flux at $\lambda < 3000$\AA\ above what is expected from conventional stellar evolution (\citealt{1969PASP...81..475C}, \citealt{1982ApJ...254..494B}). It is now generally accepted that the UV-excess (hereafter upturn) originates from hot horizontal branch (HB) stars (e.g., see review by \citealt{1999ARA&A..37..603O}; \citealt{2008ASPC..392....3Y}). Two general conditions must apply to this upturn population. First, any hot HB population giving rise to an upturn in a massive early-type galaxy cannot be metal-poor because the broadband colours and line indices of early-type galaxies (e.g. \citealt{Pastorello2014}, \citealt{1988ApJ...328..440B}) clearly show that they are dominated by a metal-rich stellar population. Second, as HB stars represent the late core Helium burning phase of low mass stars, the upturn directly probes the properties of stars that reached the zero age main sequence at relatively high redshift  and therefore provides a window to the conditions within galaxies at very early times.

\begin{figure*}
{\includegraphics[width=0.33\textwidth]{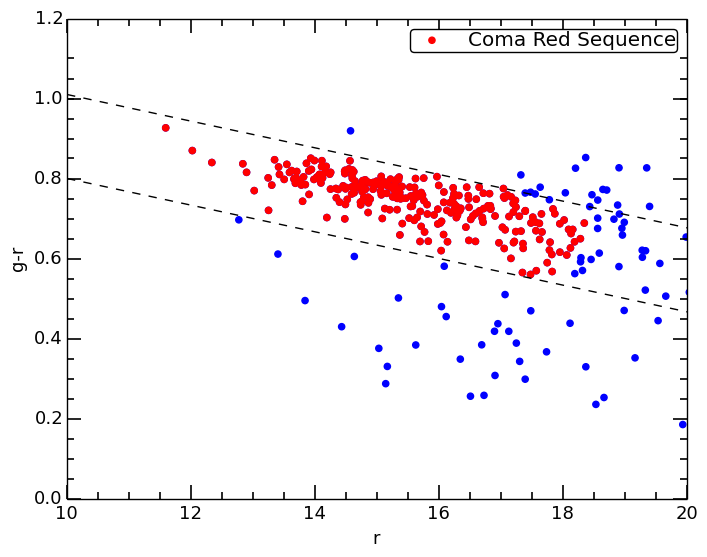}}
{\includegraphics[width=0.33\textwidth]{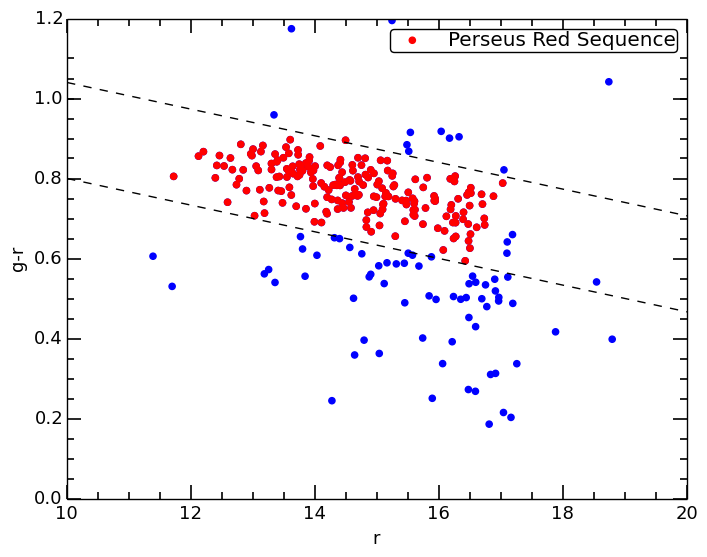}}
{\includegraphics[width=0.33\textwidth]{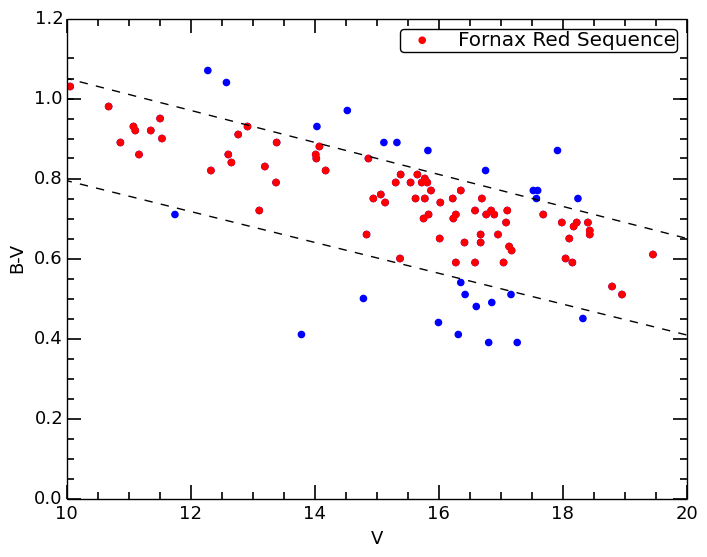}}
\caption{Optical colour-magnitude diagrams ($g-r$ or $B-V$ vs. r or $V$) for the Coma, Perseus and Fornax clusters (left to right). The red sequence is denoted by the dashed lines in each case. Galaxies used in this analysis are those red sequence objects with photometric uncertainties in their optical colour $<0.05$ magnitudes. All  colours are corrected for foreground extinction as described in the main text.}
\label{fig:redseq}
\end{figure*}

The upturn, as measured by the $FUV-V$ colour, appears to be correlated with metallicity (more properly, Mg$_2$ strength) and velocity dispersion, in the sense that more massive and metal-rich galaxies have bluer UV-optical colours (\citealt{1988ApJ...328..440B}, \citealt{2011MNRAS.414.1887B}). However, it shows very weak correlation with the Iron-abundance indices (Fe5015), suggesting that the source of the upturn is related to the $\alpha$-enriched older stellar populations (\citealt{1999ARA&A..37..603O}, \citealt{2011MNRAS.414.1887B}). \cite{2005ApJ...629L..29B} extended the earlier work by performing a cluster-wide survey for the upturn in all early-type galaxies in the Virgo cluster, going all the way down to dwarf ellipticals. They recovered the correlations between $FUV-V$ and metallicity as with \cite{1988ApJ...328..440B} in the brightest galaxies, however this correlation did not necessarily hold true for the fainter galaxies. Even though the upturn tended to get stronger with increasing mass (or luminosity) in giant ellipticals, they found that the opposite was true for dwarf ellipticals. \cite{2012MNRAS.421.2982S} recovered a similar result when looking at a large sample of optically-selected giant early type galaxies in the Coma cluster and found a reddening in the $FUV-i$ colour as the galaxy luminosity decreased (i.e. mass decreased). \cite{Schombert2016} also carried out multi-colour photometry for a large sample of local ellipticals using a combination of GALEX, SDSS, 2MASS and Spitzer data. He found that the NUV-optical colours in particular were a direct measure of the UV upturn and changed with galaxy luminosity. These colours could only be reproduced by using models with a range of blue horizontal branch populations.

One promising mechanism which produces a UV upturn is having a Helium enriched population of stars on the HB. While an increase in metallicity generally decreases the temperature of a stellar population due to increasing opacity in the stellar envelope, if the Helium abundance is sufficiently high, it acts to decrease the opacity such that the measured surface temperature of the star increases. In particular, for HB stars where the envelope is already significantly thinned, this can cause a substantial rise in the measured surface temperature of the stars compared to those with more normal Helium abundances. This allows the possibility that even in systems that have high metallicity, a significant (metal-rich) blue HB component can exist, clearly important in explaining UV upturns in early-type galaxies (\citealt{chung2017}, \citealt{norris2004}, \citealt{lee2005he}, see also review by \citealt{1999ARA&A..37..603O}).

Besides Helium enriched blue horizontal branch stars being a potential origin of the upturn, other models have also been proposed to explain this phenomenon, in particular with enhanced mass loss in metal rich stars (where these stars lose more mass as they ascend the red giant branch - \citealt{Yi1997}) and the binary evolution model (similar to the hot subdwarfs in our Galaxy - \citealt{han2007}). However, neither of these mechanisms has observational support (see below in our Discussion) and would not account for observations in local open and globular clusters.

In this paper we expand upon the work of \cite{2005ApJ...629L..29B} and \cite{2012MNRAS.421.2982S}. The approach we take is as follows: we construct the most detailed UV SED possible from the available photometry for red sequence galaxies drawn from the Coma cluster, combining each with their optical SED. We then fit the entire SED with a combination of a conventional old stellar population and a blackbody. The conventional stellar population does not include a UV-luminous post-main sequence population with which to explain the UV upturn; the additional blackbody is used to model and characterise this. The best fit temperature of this blackbody component could be interpreted as the luminosity-weighted temperature of any population that gives rise to the UV upturn. We also compare the FUV colours of the Coma galaxies with those of similar galaxies in the Fornax and Perseus clusters in order to analyse the strength of the upturn in all three. \cite{Lee2005} carried out photometry of some of the brightest galaxies in the Fornax cluster, however given the subsequent availability of much deeper GALEX data, we are able to detect a larger number of galaxies and present a more comprehensive study of this cluster.

In the following all magnitudes quoted are in the AB system, 
and the cosmology used assumes $h=0.7,\Omega_m=0.3,\Omega_\Lambda=0.7$.  We correct for Galactic extinction for Perseus galaxies in all bands using the extinction maps provided by \cite{2011ApJ...737..103S}. However, due to the large amount of extinction present particularly in the UV bands, and the possibility of there being slightly varying amounts of extinctions along different lines of sight, there is a small yet intrinsic uncertainty in the extinction-corrected UV photometry for the Perseus cluster.

\section{Data}

\subsection{Optical}

Our main focus is on the Coma cluster, but we also carry out  a comparison with the Fornax and Perseus clusters.
We select the optical cluster red sequence, as this is known to host the majority of cluster members and to consist of mainly `red and dead' E/S0 galaxies \citep{1992MNRAS.254..601B}. Data for Coma and Perseus are taken from the latest release (DR12) of the Sloan Digital Sky Survey (SDSS; \citealt{2015ApJS..219...12A,2000AJ....120.1579Y}). We select only those red sequence members with uncertainties $<0.05$ magnitudes in their relevant optical colours. We use a metric (7.5kpc diameter) aperture to measure UV-optical colours. For Fornax we use the $B$ and $V$ data from \cite{2003MNRAS.344..188K}\footnote{In this case we use their Kron (model) apertures.} and for Coma and Perseus, we convert (where necessary) the SDSS $g$ and $r$ band magnitudes to equivalent $V$ magnitude using the equations from \cite{2005AJ....130..873J}. Galaxies on the red sequence are assumed to be cluster members, as is usually the case, but we also use published catalogs to check membership (Coma: \citealt{2008A&A...490..923M,2010ApJS..191..143H} and 
Trentham et al. (priv. comm.); Fornax:
\citealt{1988AJ.....96.1520F,2000A&A...355..900D}; Perseus: \citealt{1999A&AS..139..141B}). We plot these red sequences
in Fig.~\ref{fig:redseq}.

\subsection{GALEX}
The UV images for the Fornax, Perseus and Coma clusters were obtained from archived Galaxy Evolution Explorer (GALEX - \citealt{2005ApJ...619L...1M,2007ApJS..173..682M}) data, available through the Multi-Mission Archive at the Space Telescope Science Institute (MAST). Images were obtained for both GALEX filters: FUV and NUV which have central wavelengths 1526\AA\ (range of 1344-1786\AA) and 2329\AA\ (range of 1771-2831\AA) and image resolution FWHM of $4.3''$ and $5.3''$ respectively. 

All images were retrieved as calibrated intensity maps (plus their relative $\sigma$ images) from the GalexView tool (https://galex.stsci.edu/galexview). Exposure times for Coma
images were 19ks and 30ks in the FUV band (\citealt{hammer2010}). For Perseus the exposure times in FUV were 14.9ks and 8ks (\citealt{oconnell2005}), and for Fornax we used a deep exposure of 34ks in the centre and a `ring' of 12 2.5ks exposures in the outer regions (NGS - Nearby Galaxy Survey, \citealt{gil2007}).

We measured FUV and NUV magnitudes for red sequence galaxies in these clusters through 7.5kpc (diameter) apertures, centred on the optical position. We only consider objects with $>5\sigma$ detections and verify their reality by eye. This process ensured that all red sequence Coma galaxies to 2 magnitudes below $M^*$ in the H-band (\citealt{eisenhardt2007}) were included in our sample.

\subsection{UVOT}

UVOT (Ultraviolet Optical Telescope) is an ancillary 30cm telescope on the SWIFT satellite. It has a field of view of $17' \times 17'$ with a pixel scale of 0.502$''$ and a PSF FWHM of $2.5''$. Technical descriptions of the SWIFT UVOT telescope and its photometric calibrations are given in \cite{2005SSRv..120...95R} and \cite{2008MNRAS.383..627P}. UVOT images for Coma were taken through the UVW2 (central wavelength 1928\AA), UVM2 (2246\AA), UVW1 (2600\AA) filters, as well as optical filters. We only use the UVW2 and UVW1 data in conjunction with the optical U, B and V UVOT images given that the UVM2 band largely overlaps with the GALEX NUV band and has a far shorter total exposure time in comparison.

For Coma, we were able to find images in the Swift UVOT archive from multiple programs (\citealt{brown2014}, \citealt{markwardt2005}). These covered a subset of the Coma galaxies observed with GALEX and depending upon the number and/or depth of individual exposures, resulted in a variety of effective exposure times. The deepest exposures came from overlapping pointings that targeted the centre of the cluster. Most of the UVOT-targeted galaxies ($\sim$24 out of 42) are drawn from this region and consequently have deeper ($\sim$2.7-3.4ks in UVW2 and $\sim$0.8-1.3ks in UVW1) exposure times. For the galaxies outside of this region, exposure varied between $\sim$0.8-2.7ks in UVW2 and $\sim$0.15-0.8ks in UVW1. Despite the variation in exposure times, all galaxies detected by GALEX and targeted by UVOT are detected. However while the GALEX observations sampled passive galaxies down to $M_v=-18$ (Fig. \ref{fig:coma_fornax_perseus}), the UVOT pointings only targeted one galaxy fainter than $M_v>-19$.

The images were aligned and averaged together using the \textit{imcombine} function in \textit{IRAF} to create final images with longer exposure times to perform photometry on. Similar to GALEX, photometry was carried out in all UVOT bands through 7.5kpc diameter apertures centered on the optical positions of red sequence galaxies. When combined with GALEX FUV and NUV, this creates essentially  contiguous coverage of the entire ultraviolet region of the spectrum between $\sim$1000\AA\ and $\sim$3000\AA. 

\section{Results}

\subsection{The UV upturn in Coma, Fornax and Perseus}
To analyze and compare the UV upturn in Coma, Fornax and Perseus clusters, we plot the $FUV-V$ colour, the standard measure for the strength of the UV upturn, in fixed 7.5 kpc diameter apertures of all the cluster red sequence galaxies against their absolute V band magnitudes as shown in Fig. \ref{fig:coma_fornax_perseus}. The brightest giant ellipticals tend to have the bluest colours and as such the strongest upturn. This is (as we will show) because they have hotter and more prominent extended HBs. We see a general trend of the $FUV-V$ colour getting redder with decreasing luminosity (mass). At $M_v\gtrsim-18$ however, as seen in Fornax galaxies, the $FUV-V$ trend with $V$ 
is reversed and becomes bluer with decreasing luminosity. Obviously, the lack of systems with comparatively red $FUV-V$ colours at these magnitudes is simply a reflection of the FUV magnitude limit. However, the presence of galaxies with colours significantly bluer than $FUV-V<5.5$ indicates that a significant fraction of the fainter dwarfs either have ongoing star-formation or a UV upturn population with different characteristics to those of the more luminous galaxies, perhaps unsurprising given the mass and overall metallicity for these galaxies.

So far, these observations confirm the conventional view of the upturn as shown in \cite{1988ApJ...328..440B} and \cite{2011MNRAS.414.1887B}. Note that the data for Perseus is subject to a comparatively large and therefore less certain Galactic extinction correction, and as a consequence there may be an uncertain offset in the $FUV-V$ colours between Perseus and the other clusters. However, the range in $FUV-V$ colours is largely unaffected by this and is consistent with that found in the Coma and Fornax clusters (and Virgo, \citealt{2005ApJ...629L..29B}).

Focusing only on the giant ellipticals, i.e. galaxies with $M_v\lesssim-18$, even with the brightest (most massive) galaxies tending to be somewhat bluer than their fainter counterparts, there is still significant stochastic variation of $1.5\sim2$ magnitudes in range in the $FUV-V$ colour between the majority of galaxies in all three clusters, ignoring the few outliers. Where they have bands in common, the photometric behaviour for each of the cluster populations appears consistent to the fidelity of the data, indicating that our results are most likely applicable to the general population of quiescent cluster early-type galaxies in the local Universe. However, this can only be verified with future studies involving larger sample of low redshift clusters (Ali et al. in prep).

The photometric range in UV-optical colour observed here agrees with that seen in the results of \cite{2005ApJ...629L..29B} who also found approximately 2 magnitudes of scatter in the $FUV-V$ colour in red sequence Virgo cluster galaxies. As such the variation in the UV upturn strength appears to be a ubiquitous feature present in all low redshift early-type cluster galaxies. This may be interpreted as a variation in the temperature and/or strength of an extended blue HB component in these galaxies.

If the blue colours at fainter magnitudes are due to an extended HB population in fainter galaxies rather than star-formation, these may be relatively low metallicity stars. In this case the observed trend of $FUV-V$ becoming bluer with decreasing luminosity could simply reflect the lower mean metal abundance of the stellar population of dwarf galaxies (i.e., the conventional first parameter of HB morphology). 

\subsection{The UV SEDs of Coma galaxies}

\begin{figure}
\includegraphics[width=0.5\textwidth]{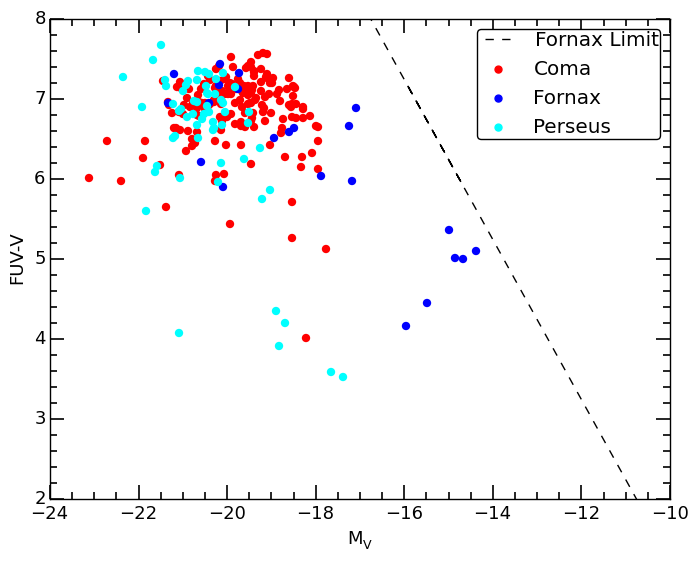}
\caption{$FUV(GALEX)-V$ vs $M_v$ colour-magnitude diagram for red sequence galaxies in Coma, Fornax and Perseus clusters. The black dashed line shows the GALEX FUV detection limit ($>5\sigma$) for Fornax. Photometric error in colour is $<0.1$ magnitudes.}
\label{fig:coma_fornax_perseus}
\end{figure}

To properly investigate the nature of the sources of the UV upturn, we would need detailed UV spectra of the galaxies in question, data that is not available at present. The broad consensus is that the UV flux is produced by hot HB stars. \cite{1997ApJ...482..685B} obtained UV spectra of some bright galaxies from Ultraviolet Imaging Telescope (UIT) and found their temperatures to be consistent with a hot HB with $T_{eff} \sim 20000$K. Imaging of the resolved sources of the UV upturn in M31 and M32 \citep{1998ApJ...504..113B} confirms their nature as stars in the extended HB.

\begin{figure}
\includegraphics[width=0.5\textwidth]{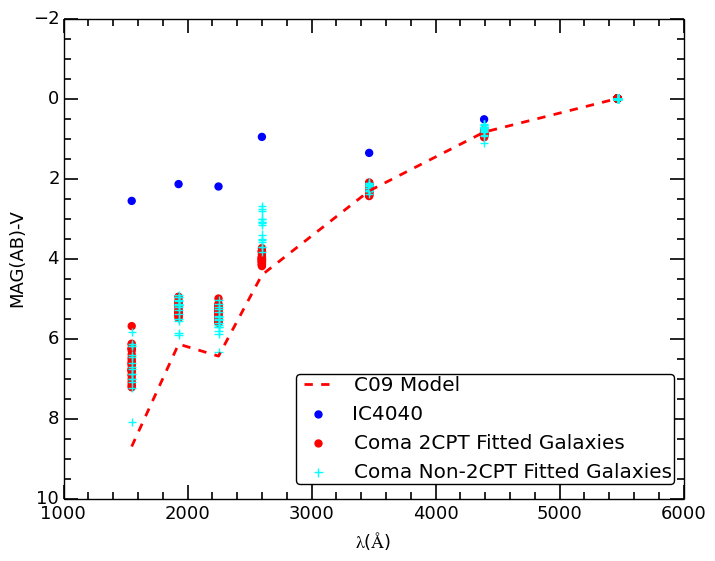}
\caption{UV to optical SEDs of Coma red sequence galaxies, IC4040 (a starforming galaxy) and an old metal-rich SSP (no starformation) from the C09 model. The colours plotted (from left to right) are $FUV-V$, $UVW2-V$, $NUV-V$, $UVW1-V$, $U-V$ and $B-V$. Photometric error in colour is $\lesssim0.1$ magnitudes.}
\label{fig:coma_sed}
\end{figure}

We have used the GALEX and UVOT data for the Coma cluster
to build a detailed spectral energy distribution for red sequence galaxies within the UVOT and GALEX images, covering between 1000 and 3000 \AA\ uniformly for these objects. From GALEX we use data from both the FUV and NUV bands and from UVOT we use data from the UVW2 and UVW1 to give us a contiguous coverage in the UV. We also make use of the UVOT U, B and V data to give us coverage in the optical all the way up to approximately 6000\AA. We plot all colours relative to V in Fig.~\ref{fig:coma_sed}.

We detected 176 confirmed early type cluster members in the GALEX bands as shown in Fig. \ref{fig:coma_fornax_perseus}, i.e. all Coma galaxies brighter than $M_v\le-18$ in the region covered by GALEX. Of these 44 were also targeted in UVOT pointings and all were detected. We then made a cut at $U-V=2.1$ (this was done in addition to the $g-r$ red sequence selection), and all galaxies with colours bluer than this limit were rejected in order to ensure that no systems with significant residual star formation were included. This left us with the 42 galaxies altogether, as shown in Fig. \ref{fig:coma_sed} (both red and cyan points in the plot combined).

\begin{figure*}
{\includegraphics[width=0.45\textwidth]{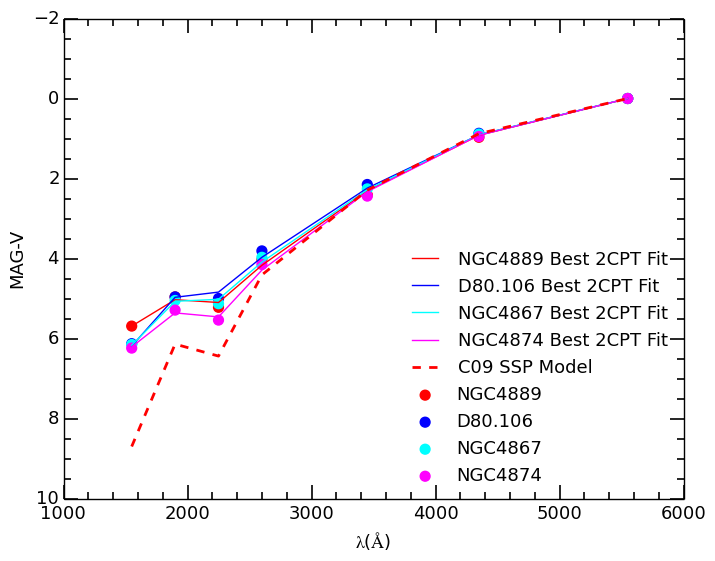}}
{\includegraphics[width=0.45\textwidth]{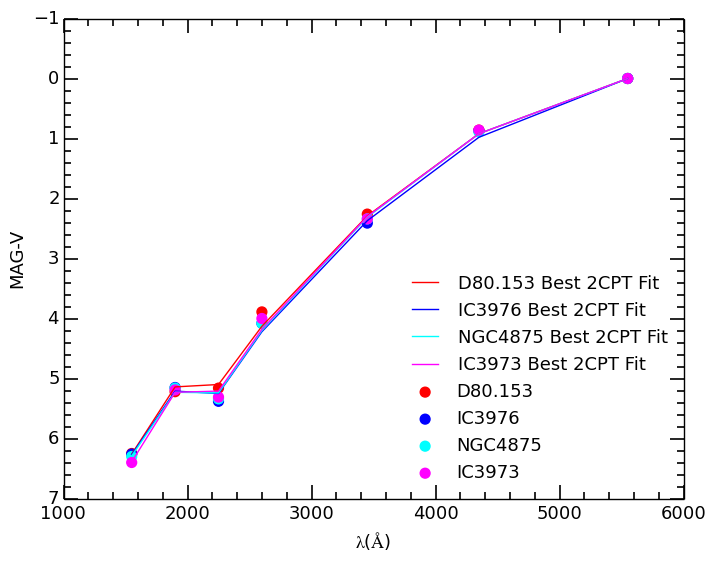}}
{\includegraphics[width=0.45\textwidth]{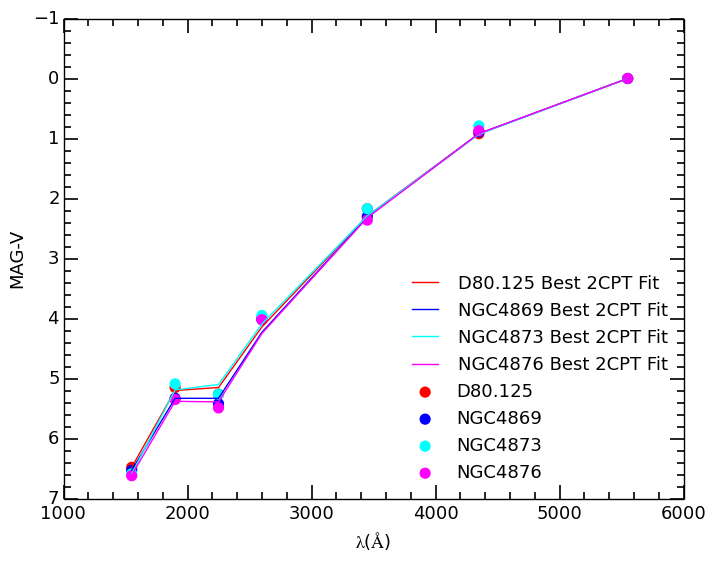}}
{\includegraphics[width=0.45\textwidth]{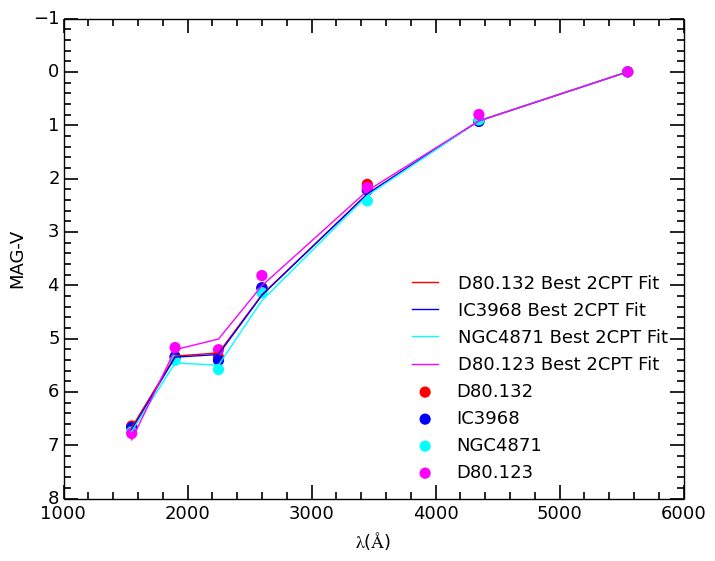}}
{\includegraphics[width=0.45\textwidth]{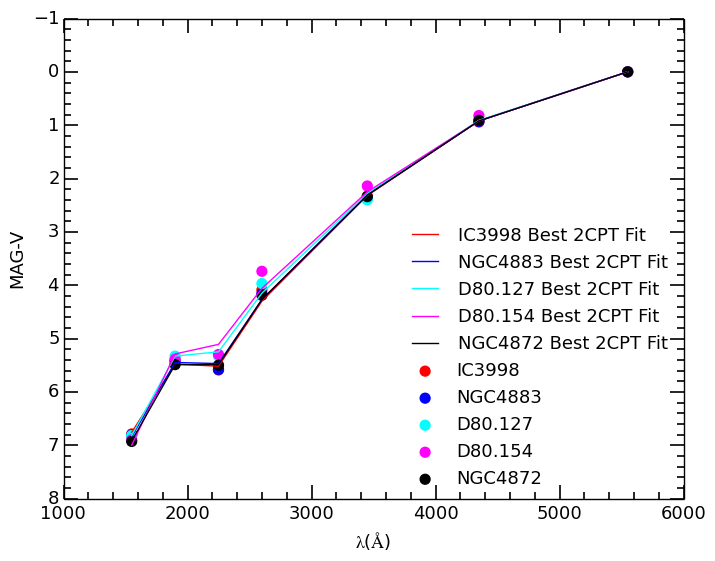}}
{\includegraphics[width=0.45\textwidth]{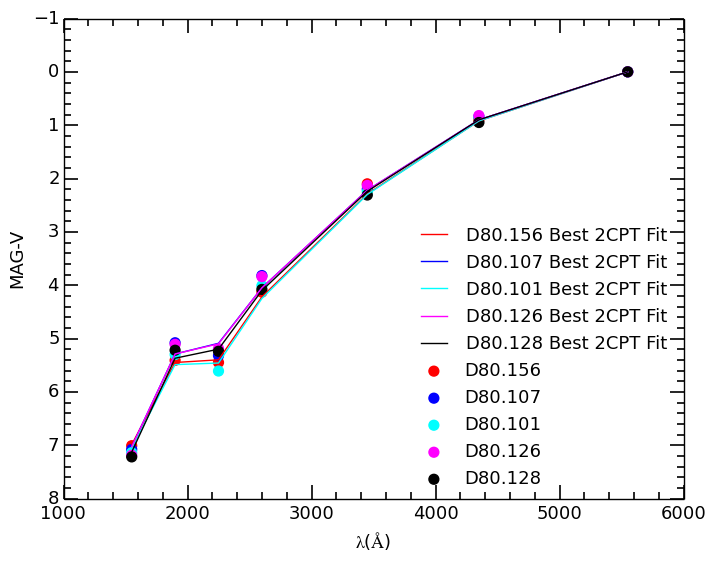}}
\caption{UV to optical SEDs of 26 Coma red sequence galaxies that were well fitted with 2 component (C09 + Blackbody) models. Galaxies are identified in the figure caption, by their NGC/IC numbers or their number in \protect\cite{dressler1980}'s catalog. Once again, the colours plotted (from left to right) are $FUV-V$, $UVW2-V$, $NUV-V$, $UVW1-V$, $U-V$ \& $B-V$. Photometric error in colour is $\lesssim0.1$ magnitudes. The dashed red line in the first panel indicates the baseline SED derived from Conroy et al. (2009) upon which the blackbody component is added to fit the individual galaxy SEDs.}
\label{fig:sed}
\end{figure*}

We also plot the SED of IC4040, a known star-forming galaxy within Coma, and the SED of a "red and dead" galaxy as given by the \cite{2009ApJ...699..486C} (C09 henceforth) model as these two SEDs represent the extreme cases of strong ongoing star formation and no star formation at all to compare with our sample of red sequence galaxies. As one would expect, the spectrum of IC4040 is much flatter going from the optical to the ultraviolet when compared to the red sequence galaxies. This is due to the IMF of a star-forming system producing stars of a range of temperatures from faint M types all the way to OB supergiants, which emit strongly at all wavelengths between optical and ultraviolet, leading to a flat spectrum. Compared to this, the upturn galaxies have a bimodal population, consisting of a majority of low temperature stars (mostly K giants by luminosity) superimposed with a relatively small but very hot sub-population of (likely) blue/extreme HB stars.

Unlike optical colours, the UV colours show considerable scatter. This is due to the different strengths, and temperatures (as we shall see) of the hot HB component. While the scatter increases towards bluer colours, as we would assume, we also find that, unexpectedly, there is large scatter in the $UVW1$ (2600 \AA)-$V$ colours, even though this filter is the reddest of the UV filters we use. We will discuss this further when fitting blackbody curves to the UV SEDs.

\subsubsection{Blackbody Fitting}

\begin{table}
\begin{filecontents*}{table.csv}
Name,Mr,BB Temp,Av,Normalisation (NUV)
NGC4889,-23.48,21000,0,340
D80.106,-19.82,13000,0,468
NGC4867,-21.14,15000,0,374
NGC4874,-23.05,19000,0,216
D80.153,-20.00,15000,0,340
IC3976,-20.59,15000,0.2,478
NGC4875,-20.47,16000,0,287
IC3973,-21.00,15000,0,298
D80.125,-19.75,14000,0,322
NGC4869,-21.48,15000,0,255
NGC4873,-20.95,12000,0.1,454
NGC4876,-20.82,15000,0,236
D80.132,-19.71,13000,0,278
IC3968,-19.78,13000,0,268
NGC4871,-21.17,15000,0,204
D80.123,-19.97,10000,0.1,500
IC3998,-20.52,15000,0,198
NGC4883,-20.96,14000,0,215
D80.127,-19.32,12000,0,291
D80.154,-19.53,10000,0.1,448
NGC4872,-20.89,13000,0,210
D80.156,-19.34,12000,0,238
D80.107,-19.87,10000,0,351
D80.101,-20.19,12000,0,220
D80.126,-19.61,10000,0,349
D80.128,-19.66,10000,0,307
\end{filecontents*}
\csvautotabular{table.csv}
\caption{Table showing the derived parameters (Blackbody temperature, Av, Normalisation) from model fitting as shown in Fig. \ref{fig:sed}. The "Normalisation" parameter gives the relative percentage normalisation of the blackbodies to the base C09 model at the GALEX NUV band (i.e. 100\% indicates that the blackbody flux is equal to the C09 flux at this wavelength.)}
\label{table:bb}
\end{table}

We now attempt to model the observed SEDs. We assume that the galaxies consist of a conventional old stellar population (with no attempt to model a hot post-MS phase) plus a secondary hot HB component. For the old SED, we choose a standard quiescent stellar population from C09, with solar metallicity and a formation redshift of $z_f=4$ (low redshift UV-optical colours change very little between $z_f=3$ and $z_f=6$ in these models). This is found to reproduce the optical colours of early-type galaxies well and produces `normal' very red UV colours ($FUV-V \sim 8.5$) that are consistent with the expectations of conventional stellar evolution models. Importantly, this SED matches the UV colours of 47 Tucanae, \citep{2012AJ....144..126D} and other metal-rich globular clusters as well as the reddest (but un-reddened) early-type galaxies with a HB nearly all confined to the red clump and with only a minor contribution from secondary populations with small He spreads. These are all old systems but do not have UV upturns.

We then model the hot HB with a single blackbody (of a given temperature) plus variable extinction following the Milky Way law \citep{1988ApJ...329L..33C}, to fit the UV SED of Coma early-type galaxies. We minimize least squares to fit the UV SEDs over a grid of reasonable values for temperature and extinction. It is important to note that the fitting is carried out in the UV part of the spectrum (i.e. GALEX FUV, UVOT UVW2, GALEX NUV, UVOT UVW1) since it is most affected by the upturn. Nevertheless, we add the contribution of this component to the SED of the old stellar population at every wavelength considered through to the V band. In any event, the actual contribution longward of 3000\AA\ is negligible. This works under the assumption that the spread in optical colours is small and the base C09 model chosen reflects the observed optical colours very well, which can be seen to be the case in Fig. \ref{fig:coma_sed}. Furthermore, the interstellar extinction in these galaxies would also need to be very small and we find this to be the case just from fitting to the UV part of the spectrum (see later).

Of the 42 total galaxies, we obtained very good fits to a single temperature blackbody (which we interpret as the mean blue HB star temperature) for 26 galaxies, from which we also derive a value for the interstellar extinction, which was generally low ($Av\lesssim0.2$). However, some galaxies appear to show excess flux in the UVW1 filter (2600 \AA), such that it is not possible to obtain a good fit (especially given the small number of points). These sources are represented by the SED points plotted in cyan in Fig. \ref{fig:coma_sed}; clearly the spread in the 2600\AA~ photometry is much wider than that for the other UV points. We have checked the date of observations and find no correlation with those galaxies giving a poor fit. The objects that are clearly discrepant in the $UVW1-V$ colour are randomly distributed in all other colours. It simply appears that there is anomalous behaviour in the UVW1 band for a subset of these galaxies. 

The galaxies with a poor fit to our models are rejected based on a least squares criterion. The distribution of this value across the sample is strongly bimodal (i.e. the good fits are fitted very well while the bad fits are significantly and clearly worse). The majority of poorly-fit galaxies have very blue UVW1-V colours. So although we do not do this, had we chosen to make a colour cut at approximately UVW1-V=3.8 (see Fig. \ref{fig:coma_sed}) and a priori rejected all galaxies with a bluer colour, we would have rejected the majority of our poorly-fit galaxies purely on this criterion.

The discrepant objects are largely (but not exclusively) found in a single UVOT $UVW1$ image (PSNJ13003230+2758411) which contains a few galaxies with a good fit. This may point to systematic problems with this particular image, though the fact that good fits were obtained with it and bad fits also occurred elsewhere means we cannot be certain about this. 

We note from Fig. \ref{fig:sed} that IC4040 also shows a comparatively high flux in this band, perhaps indicating that this may be a signature of residual star formation dominating in the UV. To explore this we created composite SEDs from a C09 model supplemented with a scaled version of the IC4040 SED to generate SEDs with star formation that acted to increase the $U-$band flux of the simulated SED by 1 to 20 \%  above that of the C09 value. Given the tightness of the range ($\sim 0.1$ mag) in $U-V$ colours of our selected objects, the models with up to a 10\% increase in $U-$band flux were plausible fits to our {\it optical} data. We then looked at the UV part of these simulated SEDs. Only the model with the smallest component of star formation (1\%) gave a plausible fit with the C09 plus BB model, even then it is far redder in UV-optical colours than all of our well-fitted galaxies - the star forming component only increases the UV brightness relative to C09 alone by a few tenths in magnitude.  All others fail largely because of their NUV-UVW1 colour being too steep for any C09 plus BB model. Consequently, from this exercise it is clear that star formation models do not generate the UV upturns we see in the galaxies with good fits. Any one colour could be fit, but not the entire UV-optical SED.

We also attempted to fit the SEDs of our poorly fit (C09+BB) galaxies with the C09+RSF models, but fail to get good fits to their SEDs. However, we cannot rule out the possibility that this subset of galaxies has UV emission due to some residual star formation which could be accounted for by a more sophisticated residual star forming (or upturn+RSF) models.

Furthermore, we also note that HST WFC3/F336W imaging data for red sequence galaxies in Abell~1689 (which is close to rest-frame $\sim$2600\AA) does not show the abnormally large spread in equivalent colour as seen in the UVOT data for Coma (Ali et al., in prep), indicating that this is likely an issue with some of the UVW1 data and not a genuine feature of some UV SEDs. As we cannot explain this discrepancy and it appears to be a problem with a subset of a single band's photometry, we exclude the affected objects from further discussion.

There is no intrinsic reason why the best blackbody fits could not contribute significant flux in the optical bands because the fit is not constrained there. The fact that the blackbody fits never do this strongly supports the idea that the UV is dominated by the component modelled here as a blackbody while the optical is completely dominated by the conventional old metal-rich stellar population. As a consistency check we have performed the entire model fitting process for the whole spectrum (UV and optical points), allowing both blackbody and C09 components to vary across the wavelength range and find that although there are some minor changes to the individual results obtained for each galaxy, the overall trends and conclusions remain intact.

\begin{figure}
\includegraphics[width=0.45\textwidth, height=0.3\textheight]{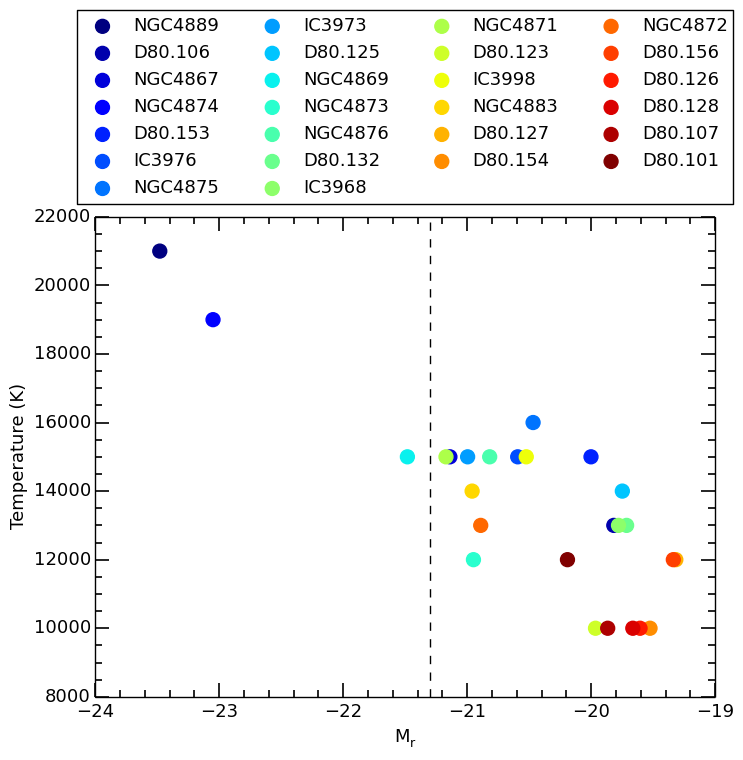}\\
\includegraphics[width=0.49\textwidth, height=0.27\textheight]{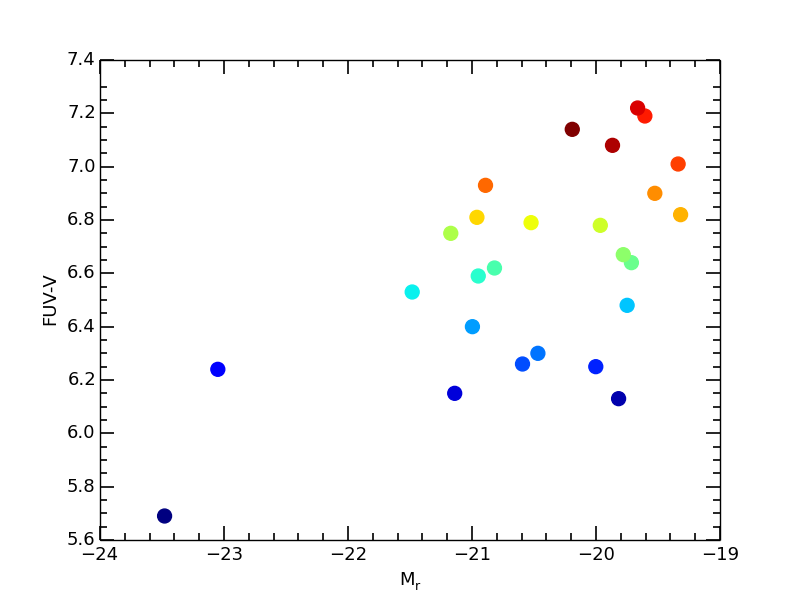}\\
\includegraphics[width=0.49\textwidth, height=0.27\textheight]{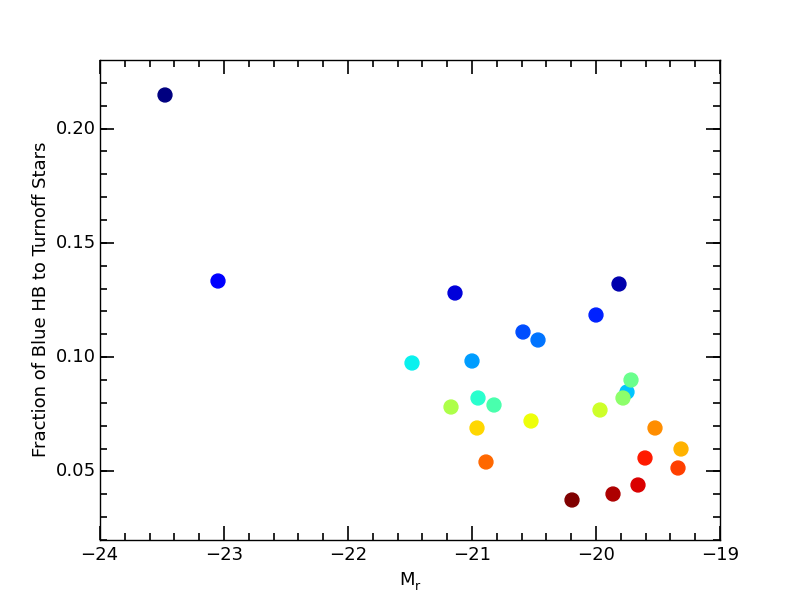}
\caption{Top: The best fit blackbody temperature vs. $M_r$ for galaxies in Coma. There is a clear trend of higher $T_{eff}$ with increasing luminosity (and hence increasing mass and metallicity).  Dashed line shows $M^*$ for Coma (\protect\citealt{eisenhardt2007}). Typical errors in temperature are of the order $\pm1000$K, the quantisation we use in our fitting process. Middle: this shows $FUV-V$ vs. $M_r$ for these same galaxies, showing how the increasing temperature drives the bluer colours. Photometric error in colour is $\lesssim0.1$ magnitudes. Bottom: The  estimate of the ratio of blue HB stars to turnoff stars within each galaxy as a function of $M_r$ for these objects.}
\label{fig:temp_mr}
\end{figure}

All major derived parameters for the good fit galaxies are given in table \ref{table:bb}. Fig. \ref{fig:sed} shows the galaxies with their corresponding best fit models. In the first panel we also show the underlying C09 model showing the contribution of the purely old population, clearly demonstrating that the extra blackbody component only contributes to the UV part of the SED. The fitted blackbody component usually has a temperature between 10,000 and $\sim21,000$K, consistent with the earlier UIT observations of \cite{1997ApJ...482..685B} and the indices derived by \cite{2016MNRAS.461..766L} for massive BOSS galaxies. We plot the derived temperature vs. $M_r$ (a good proxy for stellar mass) in Fig.~\ref{fig:temp_mr} and we find a strong correlation between temperature and mass (and therefore metal abundance), in the sense that more massive (and metal rich) galaxies have hotter (and stronger) HB components. This also recovers the $FUV-V$ vs. $V$ anti-correlation, as observed by \cite{1988ApJ...328..440B} and \cite{2011MNRAS.414.1887B}. We note here that since the hot component does not have a single 
temperature for all galaxies, one cannot use a single UV-optical colour, e.g. $FUV-V$ to measure the upturn in a uniform manner and classify sources as in \cite{2005ApJ...619L.111Y,2011ApJS..195...22Y}, a point also made by 
\cite{2012MNRAS.421.2982S}. 

\cite{nelan2005} and \cite{thomas2005, thomas2010} demonstrated through a spectroscopic analysis that both metallicity and age were correlated with the velocity dispersion and therefore mass of early-types, with \cite{price2011} demonstrating similar behaviour in Coma early-types. The variation in the 
$FUV-V$ colour of the C09 models with age beyond 8~Gyrs does not change their contribution to the UV spectrum, therefore validating our decision to use a single age model for all of our fits to the optical photometry regardless of the luminosity. Similarly, for our $g-r$ selection, metallicity is expected to vary between $1-2$\(Z_\odot\). However, as a test of consistency, we duplicated out fitting procedure using a C09 model with $Z=0.56$\(Z_\odot\), in particular to check whether such a sub-solar model yields a better fit for the lowest luminosity galaxies in our sample. We find that for all such galaxies, using a base sub-solar metallicity model gives worse fits to the optical colours than using a solar metallicity model. Furthermore, the most reasonable fits of the sub-solar metallicity models required very large values of interstellar extinction ($A_v\sim0.7$), which are clearly unphysical for these systems. Conversely, for the most luminous galaxies in our sample - which are likely to be of super-solar metallicity - a base solar metallicity C09 model is already overestimating the contribution from the underlying conventional stellar population, given that the super-solar metallicity models have redder UV-optical colours. 

To test the effect of a super-solar metallicity model, we repeated our entire fitting procedure using a base C09 SSP with $Z=1.78$\(Z_\odot\) (the next highest Z available after $Z$=\(Z_\odot\)), likely to be appropriate for the top 50$\%$ of the brightest galaxies in our sample. We find that the best fit temperatures decreased by on average $\sim1000K$, but the blackbody normalisations increased significantly between $\sim200-500\%$ to compensate for the combination of a decrease in blackbody temperature and the underlying super-solar SSP being redder in the UV. However, despite the change in temperatures and normalisations when using a base super-solar metallicity model, the overall trend seen in Fig. \ref{fig:temp_mr} (top) still remained intact.

Given the high metallicity and old ages (and therefore the restricted range of both) for the conventional stellar populations in all of these galaxies, neither their metallicity nor age can therefore be driving the correlations seen in Fig. \ref{fig:temp_mr}. Note that we are not saying that there is no correlation between overall age and strength of UV upturn in the actual systems. We are merely stating that the C09 models include no component that specifically attempts to replicate anything that may give rise to a UV upturn, unlike e.g. the \cite{Yi1999} models. The parameters of the blackbody components that we use to fit the upturn may correlate with overall age but we do not attempt to make this explicit.

\section{Discussion}

\subsection{Hot HB stars as the source of the UV upturn}

We have measured vacuum UV colours and the UV SEDs for quiescent early-type galaxies in three low redshift clusters, reaching to the $L^*$ point 
and somewhat beyond. Across this entire luminosity (mass) range we find evidence for a UV-bright stellar population that we relate to the classical 
upturn (\citealt{1999ARA&A..37..603O}). Reflecting the scatter in $FUV-V$, all of the galaxies undergoing our fitting 
procedure require an extra blackbody component on top of the C09 SED model to account for their UV SED. Moreover, the temperature of this component 
seems to increase with optical luminosity, albeit with some scatter. The need for this component is already clear in the scatter of $FUV-V$ colour 
in the absence of similar scatter in the optical colours.

Residual star formation does not produce an SED consistent with our observations (cf. our discussion of IC4040 above). The most likely source of 
the excess UV flux is a hot HB component, as earlier argued by \cite{1990ApJ...364...35G,1997ApJ...482..685B,1999ARA&A..37..603O}. While blue 
HBs are commonly observed in metal-poor star clusters it is very unlikely that our observations can be explained in this fashion. Our galaxies 
are largely metal rich and their HB should largely coincide with the red clump at temperatures of around 4000-5000K: too cool to provide the 
observed UV emission in the upturn galaxies. Given the small fraction of the lifetime that stars live on the HB, it would require a significant 
low metallicity main sequence population to exist within these galaxies, which is explicitly ruled out by many studies determining the 
metallicities of such galaxies. The explanation of these UV upturn galaxies lies in the understanding gleaned from observations of the resolved 
stellar populations in high metallicity open clusters and so-called "second parameter" globular clusters (\citealt{2009ASSP....7..175C,
2017MNRAS.464..713P}).

The most likely mechanism to explain the presence of hot HB stars in comparatively metal-rich systems (by globular cluster standards, i.e., 
where the HB should not extend beyond the RR Lyra gap according to canonical models) involves Helium enrichment. In Milky Way globular clusters 
such as $\omega$ Centauri and NGC2808, multiple main sequences are found to correspond with multi-modal blue HBs and spectroscopic evidence 
shows that the bluer stars are more metal rich than the red ones (\citealt{2005ApJ...621..777P,2007ApJ...661L..53P}, \citealt{2010ApJ...720L..41B}). 
The only way for this to be possible is for the bluer stellar population to be enriched in Helium relative to the redder population. There is 
indirect evidence of this in the selective abundance patterns of the proton-capture elements, which indicates processing through the Mg-Al chains 
during hot bottom burning, a process that may also dredge up Helium to the surfaces (e.g, see \citealt{2012A&ARv..20...50G,2016EAS....80..177C} 
for reviews). The He abundance may be as high as Y=0.43 for the most extreme population in NGC2808 \citep{2015ApJ...808...51M}. In other words 
the anomalously hot horizontal branch stars in Galactic globular clusters arise from a population with significantly enhanced Helium abundance.

Some of the more metal rich globular clusters in the bulge, such as NGC6388 and NGC6441 (which have [Fe/H] $\sim -0.5$, at the high metallicity 
tail of the distribution in the Galaxy), also have extended HBs. \cite{2017MNRAS.465.1046T} show that these stars have temperatures of up to 
14,000K for a $Y$ of 0.38. A specific Galactic system that provides a possible model for a Helium enriched high metallicity and old population is the open 
cluster NGC6791 (cf. \citealt{1995ApJ...442..105D} for a similar remark). In many ways this resembles a miniature version of elliptical galaxies.
\cite{2012ApJ...749...35B} demonstrate that it has a single stellar population with an age of 9 Gyr and [Fe/H] of +0.3. It is also considerably $\alpha$ 
element enhanced and has $Y=0.30\pm0.04$ (\citealt{Linden2017}, though see \citealt{boesgaard2015}). Its HB contains a significant 
blue population and its integrated SED is essentially indistinguishable from strong UV upturn elliptical galaxies. \cite{2012ApJ...749...35B} 
argue, given this similarity, that Helium enriched horizontal branch stars are therefore an extremely strong candidate to explain UV upturns 
in elliptical galaxies. Despite it being a single and possibly unusual object (but cf. Berkeley 17, \citealt{bragaglia2006}), NGC6791 represents the
{\it only} resolved stellar counterpart to the sources of the upturn in external galaxies and therefore remains the best heuristic model with which to interpret this UV excess in galaxies.

While He-enhanced horizontal branches have been directly observed, there are other theoretical possibilities for producing a UV upturn population. 
These include scenarios where there is extra mass loss in the red giant branch (\citealt{2008IAUS..252..261Y}, \citealt{Yi1999}), possibly due to 
binary interactions \citep{2007MNRAS.380.1098H,2017ApJ...841L..10C}.  However, in this case the degree of mass loss or the close binary fraction 
would somehow (e.g. via metallicity, \citealt{Yi1999}) have to depend strongly on galaxy mass and neither explanation would account 
for the observations in globular clusters, where the blue HB segments are clearly related, by chemical tagging, to the multiple main sequences 
\citep{2015ApJ...808...51M}. This would be in contrast with recent observations by \cite{Badenes2017}, where the binary fraction 
{\it decreases} with increasing metallicity, opposite to what would be needed to explain the observed trends. Moreover, there is 
no observational evidence for enhanced mass loss on the RGB  depending on metal abundance (e.g. see \citealt{salaris2016}, 
\citealt{miglio2012}). In any event, most of the mass loss would  be expected to happen at the AGB, i.e. after 
the horizontal branch phase.

All this points to the presence of a He-rich subpopulation in these galaxies producing hot HB stars at the end of their lifetime. Given that stars 
spend a small fraction of their lifetimes in the HB ($\sim1\%$), they represent a much larger number of He enhanced stars within the galaxy.

\subsection{Implications for galaxy formation}

Given our blackbody fits to the UV upturns, we can then use UV photometry of individual hot HB stars in local globular and open clusters to provide estimates of the properties of the UV upturn populations.  First, given the estimated temperature of our BB fit for a given galaxy, we can estimate a corresponding value of $Y$ (since higher He abundance leads to hotter HB stars). Second, from the observed absolute luminosities of local hot HB stars (e.g. from \citealt{2002ASPC..265..261S}; \citealt{recio2006}), we can deduce the number of such stars needed to account for the galaxy's FUV output. 

To relate this to a global population fraction we adopt a HB lifetime of $\sim 100$ Myr and $\sim 1$ Gyr for stars to evolve from the main sequence turnoff to the Helium flash. To estimate the fraction of blue HB stars (i.e. the He-rich population), via the initial mass function, we need to assume an age for the `normal' and He-rich populations.  As a guide to the age since formation to use for our stellar populations, we take the old, metal-rich open cluster NGC6791 as a suitable model. This requires a Helium enhanced population with $Y=0.34$ to explain its blue HB and consequent UV upturn, which is very similar in colour to those of our typical galaxies.  We
then use the YEPS spectrophotometric models for He-enhanced stellar populations (\citealt{chung2017}), with $Y=0.34$, to derive an age of $12$ Gyr
for these stars to match the observed UV colours and temperatures. That is, with this level of He enrichment, it takes $\sim 12$~Gyr for the upturn to become apparent in the YEPS models' simulated colours \citep[see also][]{Tantalo1996}. We can therefore determine the relevant masses for the turn-off stars, etc., in order to calculate the relative numbers of HB and total stars from the IMF.

In fact, if we assume that both the He-enriched and 'normal' populations are formed in effectively simultaneous coeval bursts, we should expect that the main sequence lifetimes of the He-rich population stars are shorter than those of the normal population \citep{chantereau2015}. Consequently, at any given time, the He-rich stars reaching the HB should be of lower mass than those of the other population. If we assume that HB luminosities are largely independent of mass \citep[as observed in NGC2808,][]{2007ApJ...661L..53P}, then as time goes on, given a normal IMF, an increasing number of stars enter the blue HB. If this is not matched by a shorter HB lifetime, the overall UV luminosity contributed by the He-rich HB population will increase. Thus our base estimate of the fraction of He-rich stars, from above, will be an {\it over}estimate.

However, we note that the YEPS He-enhanced SSP models imply a constant UV luminosity with time once a blue HB is formed, rather than one where the UV luminosity increases. This may be because, in their models, the time spent on the HB decreases with stellar mass, compensating for the increasing rate of stars entering the blue HB with time due to the slope of the IMF. If this is the case then we should expect our upper limits to be close to the real values.

A full treatment of this would require a detailed study of the relevant isochrones and is beyond the scope of this work. However, to progress we can make the (incorrect) assumption  that the MS lifetimes of the two  populations are the same (leading to the same mass stars joining the HB at the same time) and thereby determine an upper limit to the fraction of stars in the He-rich population for each of our galaxies. By then making an estimate of the correction to the relevant mass of the He-rich population and assuming a standard slope for the IMF, we can estimate a correction to this upper limit, thereby indicating whether the derived  upper limits are likely to be close to  the true values. 

Given the above, we consider ages since formation of both stellar populations to be $12$ Gyr. Not only is this  consistent  with the indication given by NGC6791, it also agrees  with spectrophotometric and spectroscopic measurements on early-type galaxies with masses similar to those used here, where red sequence galaxies are  found to evolve passively and with no increase in  colour spread at least to $z\sim2$, and probably higher, with quiescent galaxies observed to $z\sim4$, including in the Dark Energy Survey data (C. Maraston, priv. comm.) and elsewhere (e.g. \citealt{2017Natur.544...71G}).This points to high redshifts of star formation and no significant further episodes at later times (see also \citealt{thomas2010}). Assuming these ages,
we find that a range of between  4 and 20\% of stars in our sample galaxies evolve  onto the blue HB. This is  a similar fraction as that suggested by \cite{1995ApJ...442..105D}
and observed in some local stellar clusters (see below), and also as estimated by \cite{2016MNRAS.461..766L} for massive BOSS galaxies. The fractions and
temperatures of these stars appear to depend on the luminosities of their parent galaxies, suggesting that the origin of the hot HB population is
related to the internal chemical evolution of galaxies.

Similar behaviour is seen in a minority of local 
globular and open clusters which have a wide spread in Helium abundances within their stellar populations (with NGC2808 and $\omega$ Centauri being 
the prime examples). In these cases 10--30\% of stars seem to have $Y \sim$ 0.34--0.40 \citep{2007ApJ...661L..53P,2012ApJ...749...35B,2017MNRAS.465.1046T}. 
Note that a significant fraction of the population may also be more He rich than the cosmological value but do not produce hot HB stars because of 
their high metallicities; this appears to be the case for the objects studied by \cite{2017MNRAS.465.1046T}, where 2/3 of the stars may have $Y > 0.24$.

Individual HB stellar temperatures appear to correlate with  Helium abundance e.g., for NGC2808 $T_{eff}$ of nearly 40,000K for stars with $Y=0.43$. Most of our galaxies are fitted with $T_{eff} \sim 15,000$K indicating that any He-enhanced population within them have lower $Y$  than this, similar to that of NGC6791. This allows us to assess any correction required to the fraction of stars in this He-enhanced population due to the assumption of equal stellar mass used in the estimation of the upper limit. In \cite{chantereau2015} a $Y=0.4$ star with  0.6 M$_\odot$ has a main-sequence lifetime of  11.1 Gyrs, compared to 11.9 Gyr for a 0.8M$_\odot$
counterpart with cosmological He abundance.  If we assume that stars of these masses reach the HB at the same time, then for a typical IMF slope, this implies twice as many H-enriched stars than if we assumed (as above) that the MS lifetime was independent of Y. Consequently, assuming that the increased rate of entry onto the HB by He-enriched stars is not matched by a decrease in their HB lifetime {\it and} that HB luminosity does not vary overly with stellar mass, this difference implies that the true fraction of stars with enhanced He abundances should be within a factor $\sim 2$ of our simple (upper limit) estimate. Given that all except the most massive of our galaxies are likely to have $Y<0.4$, even this factor $\sim 2$ is likely to be itself an upper limit. One final {\it caveat} on this estimate is that even though we might expect a $Y-$related variation in  turn-off  mass and main sequence lifetime, no difference in the MS turnoff mass has been reported for any of the stellar populations with varying $Y$ in NGC2808 (\citealt{2015ApJ...808...51M}).

\subsubsection{Effect on early evolution}
Assuming our blackbody component can be interpreted as a Helium enhanced population, this population appears to be more enriched and more abundant as a function of galaxy mass. As galaxy mass and metallicity are in general thought to be correlated in early type galaxies, this may suggest a relation between Y and Z, unsurprising given the predictions of many models of stellar nucleosynthesis. This suggests that these objects originate from the normal process of galaxy chemical evolution and are formed in situ. We know from \cite{2011MNRAS.414.3410C} that the hot HB component is centrally concentrated, as expected if the He enrichment comes from the same processes of chemical evolution. Given that our observations mirror those of \cite{2012ApJ...749...35B}, our results therefore imply a rapid enrichment for a large mass of stars in situ, and therefore rapid formation of massive halos at high redshifts (in agreement with e.g., \citealt{2017Natur.544...71G,jorgensen2017}).

Using the same assumptions as those in \cite{chung2017} and \cite{2012ApJ...749...35B}, if we take $\sim 12$~Gyr as a typical age for our stellar populations (as above) our data allow us to set a lower limit to the stellar mass of these 
galaxies at $z \sim 4$. Assuming 4-20\% of the stars to be sufficiently He rich to evolve to the hot HB (Fig. \ref{fig:temp_mr}, bottom), a $L^*$ galaxy must have a stellar  mass of at least 
$0.3-0.8 \times 10^{10}$ $M_{\odot}$ in place at $z \sim 4$ or earlier, with all the caveats noted above. This in turn  implies that the bulk of the stellar population may be even older. Varying the assumed age of the stellar population by 2 Gyrs only changes the fraction of He rich stars by $\sim 1.5$\%. In any event, there needs to be an earlier population to produce the extra Helium. The only way this argument may be incorrect is if the $Y$ value of this population is even more extreme, in which case the stars can reach the HB in a significantly shorter period of time. Given the abundance yields of any prior population that gives rise to the enhanced He fractions of the hot HB stars, it is likely that this earlier population must be an order of magnitude more massive than the values estimated above \citep{2016MNRAS.458.2122D}. This would imply that nearly all of the stellar mass observed today in these galaxies is in place at high redshift. We note that even if our assumption of typical $Y\sim 0.34$ and therefore 12 Gyrs as an age for stellar populations of our typical galaxies is incorrect, the lack of evidence for appreciable star formation at $z<2$ (\citealt{jorgensen2017} find formation redshifts for cluster galaxies between $2<z<6$ depending on the diagnostics used) limits their age to $>10$ Gyrs (requiring higher $Y$). Even then, similar arguments can be made as to how much stellar mass needs to be present by this time.

There have been alternative models invoked to explain an increased Helium abundance in galaxies, for example those involving Helium sedimentation in galaxy-cluster scale halos \citep{Peng2009}. Here, we would note that our results seem to run counter to the predictions of Helium sedimentation. Our most massive galaxies appear to have the highest Helium enhancement and likely contain the oldest stellar populations. Helium sedimentation acts to increase the He enhancement with increasing time and over timescales of billions of year.

If a helium enhanced sub-population is not accounted for, it is possible that galaxy properties derived from photometry and spectroscopy may be misleading or misinterpreted. Interpretation of spectral indices, particularly those involving Balmer lines may be incorrect if the Helium abundance and the presence of a hot horizontal branch is not correctly accounted for \citep{1995A&A...302..718D,2000MNRAS.316L...9D,2011MNRAS.412.2445P} and so derived quantities such as ages and metallicities of old stellar populations might appear self consistent, yet be inaccurate.The YEPS photometric models \citep{chung2017} show the required behaviour for late-time evolution, showing UV upturns, but imply that for reasonable metallicities, most broad band optical colours are unaffected by varying He abundance over the range modelled at intermediate ages. However, stellar populations younger than 1~Gyr are not presented  and so the effect on the colours of {\it e.g.} Lyman break galaxies and other young systems is unclear. 

\section{Conclusions}

We have shown that the UV upturn is a common phenomenon in quiescent early-type galaxies to luminosities well below the $L^*$ point. All galaxies appear to show at least some fraction of a UV-bright population. Characterising  this upturn with a blackbody component,  the temperature and mass of this population appears to vary with the mass (and therefore metallicity) of the galaxy, in the sense that brighter and more metal-rich systems have the hottest and more luminous UV fluxes. We interpret this, in common with other studies, as evidence for the presence of hot HB stars in these galaxies. This interpretation of our results indicates that the more massive galaxies have the most extreme HBs.  An explanation of this consistent with the findings of local stellar populations (e.g. in star clusters) is that these stellar populations are systematically enriched in Helium. If this is the case, then, noting the caveats discussed in section 4.2, we can estimate that up to 4-20\% of the turnoff stars (galaxy-dependent) must be so enriched as to evolve on to the blue HB, with $Y$ of about 0.34 to explain the  observations of our typical galaxies. 

Interpreting our results via the recent YEPS stellar population synthesis models of \cite{chung2017} implies that at least $\sim0.5 \times 10^{10}$ $M_{\odot}$ from the He-enriched sub-population alone needs to be in place, certainly by $z=2$ and by $z\sim4$ if our assumption of $Y\sim 0.34$ is reasonable. This interpretation implies that there is an upper limit to the redshift at which upturns are easily detectable, given the main sequence lifetimes of the He-enriched stars that go on to become hot HB stars. This limit may itself be dependent on galaxy mass through the correlation between this quantity and the current-day UV upturn properties. We will explore this in future papers. 

\section{Acknowledgments}
SSA is funded by an STFC PhD studentship. SSA, MNB and SP wish to thank the University of Turku and FINCA for their hospitality and local funding during their visits while working on this project. 
Funding for SDSS-III has been provided by the Alfred P. Sloan Foundation, the Participating Institutions, the National Science Foundation, and the U.S. Department of Energy Office of Science. The SDSS-III web site is http://www.sdss3.org/. This work was based on observations made with the NASA Galaxy Evolution Explorer. GALEX is operated for NASA by the California Institute of Technology under NASA contract NAS5-98034.  We acknowledge the use of public data from the Swift data archive.

We thank the referee Russell Smith for an insightful and thorough review which helped to improve this paper.

\bibliographystyle{yahapj}
\bibliography{references}

\begin{thebibliography}{}
\providecommand\natexlab[1]{#1}
\providecommand\JournalTitle[1]{#1}

\bibitem[{{Alam} {et~al.}(2015){Alam}, {Albareti}, {Allende Prieto}, {Anders},
  {Anderson}, {Anderton}, {Andrews}, {Armengaud}, {Aubourg}, {Bailey}, \&
  et~al.}]{2015ApJS..219...12A}
{Alam}, S., {Albareti}, F.~D., {Allende Prieto}, C., {et~al.} 2015,
  \href{http://dx.doi.org/10.1088/0067-0049/219/1/12}{\JournalTitle{\apjs},
  219, 12}

\bibitem[{{Badenes} {et~al.}(2017){Badenes}, {Mazzola}, {Thompson}, {Covey},
  {Freeman}, {Walker}, {Moe}, {Troup}, {Nidever}, {Allende Prieto}, {Andrews},
  {Barb{\'a}}, {Beers}, {Bovy}, {Carlberg}, {De Lee}, {Johnson}, {Lewis},
  {Majewski}, {Pinsonneault}, {Sobeck}, {Stassun}, \& {Zasowski}}]{Badenes2017}
{Badenes}, C., {Mazzola}, C., {Thompson}, T.~A., {et~al.} 2017,
  \JournalTitle{ArXiv e-prints},
  \href{http://arxiv.org/abs/1711.00660}{{\sffamily arXiv:1711.00660
  [astro-ph.SR]}}

\bibitem[{{Bertola} {et~al.}(1982){Bertola}, {Capaccioli}, \&
  {Oke}}]{1982ApJ...254..494B}
{Bertola}, F., {Capaccioli}, M., \& {Oke}, J.~B. 1982,
  \href{http://dx.doi.org/10.1086/159758}{\JournalTitle{\apj}, 254, 494}

\bibitem[{{Boesgaard} {et~al.}(2015){Boesgaard}, {Lum}, \&
  {Deliyannis}}]{boesgaard2015}
{Boesgaard}, A.~M., {Lum}, M.~G., \& {Deliyannis}, C.~P. 2015,
  \href{http://dx.doi.org/10.1088/0004-637X/799/2/202}{\JournalTitle{\apj},
  799, 202}

\bibitem[{{Boselli} {et~al.}(2005){Boselli}, {Cortese}, {Deharveng}, {Gavazzi},
  {Yi}, {Gil de Paz}, {Seibert}, {Boissier}, {Donas}, {Lee}, {Madore},
  {Martin}, {Rich}, \& {Sohn}}]{2005ApJ...629L..29B}
{Boselli}, A., {Cortese}, L., {Deharveng}, J.~M., {et~al.} 2005,
  \href{http://dx.doi.org/10.1086/444534}{\JournalTitle{\apjl}, 629, L29}

\bibitem[{{Bower} {et~al.}(1992){Bower}, {Lucey}, \&
  {Ellis}}]{1992MNRAS.254..601B}
{Bower}, R.~G., {Lucey}, J.~R., \& {Ellis}, R.~S. 1992,
  \href{http://dx.doi.org/10.1093/mnras/254.4.601}{\JournalTitle{\mnras}, 254,
  601}

\bibitem[{{Bragaglia} {et~al.}(2006){Bragaglia}, {Tosi}, {Andreuzzi}, \&
  {Marconi}}]{bragaglia2006}
{Bragaglia}, A., {Tosi}, M., {Andreuzzi}, G., \& {Marconi}, G. 2006,
  \href{http://dx.doi.org/10.1111/j.1365-2966.2006.10266.x}{\JournalTitle{\mnras},
  368, 1971}

\bibitem[{{Bragaglia} {et~al.}(2010){Bragaglia}, {Carretta}, {Gratton},
  {Lucatello}, {Milone}, {Piotto}, {D'Orazi}, {Cassisi}, {Sneden}, \&
  {Bedin}}]{2010ApJ...720L..41B}
{Bragaglia}, A., {Carretta}, E., {Gratton}, R.~G., {et~al.} 2010,
  \href{http://dx.doi.org/10.1088/2041-8205/720/1/L41}{\JournalTitle{\apjl},
  720, L41}

\bibitem[{{Brown} {et~al.}(2014){Brown}, {Breeveld}, {Holland}, {Kuin}, \&
  {Pritchard}}]{brown2014}
{Brown}, P.~J., {Breeveld}, A.~A., {Holland}, S., {Kuin}, P., \& {Pritchard},
  T. 2014,
  \href{http://dx.doi.org/10.1007/s10509-014-2059-8}{\JournalTitle{\apss}, 354,
  89}

\bibitem[{{Brown} {et~al.}(1997){Brown}, {Ferguson}, {Davidsen}, \&
  {Dorman}}]{1997ApJ...482..685B}
{Brown}, T.~M., {Ferguson}, H.~C., {Davidsen}, A.~F., \& {Dorman}, B. 1997,
  \href{http://dx.doi.org/10.1086/304187}{\JournalTitle{\apj}, 482, 685}

\bibitem[{{Brown} {et~al.}(1998){Brown}, {Ferguson}, {Stanford}, \&
  {Deharveng}}]{1998ApJ...504..113B}
{Brown}, T.~M., {Ferguson}, H.~C., {Stanford}, S.~A., \& {Deharveng}, J.-M.
  1998, \href{http://dx.doi.org/10.1086/306079}{\JournalTitle{\apj}, 504, 113}

\bibitem[{{Brunzendorf} \& {Meusinger}(1999)}]{1999A&AS..139..141B}
{Brunzendorf}, J., \& {Meusinger}, H. 1999,
  \href{http://dx.doi.org/10.1051/aas:1999111}{\JournalTitle{\aaps}, 139, 141}

\bibitem[{{Bureau} {et~al.}(2011){Bureau}, {Jeong}, {Yi}, {Schawinski},
  {Houghton}, {Davies}, {Bacon}, {Cappellari}, {de Zeeuw}, {Emsellem},
  {Falc{\'o}n-Barroso}, {Krajnovi{\'c}}, {Kuntschner}, {McDermid}, {Peletier},
  {Sarzi}, {Sohn}, {Thomas}, {van den Bosch}, \& {van de
  Ven}}]{2011MNRAS.414.1887B}
{Bureau}, M., {Jeong}, H., {Yi}, S.~K., {et~al.} 2011,
  \href{http://dx.doi.org/10.1111/j.1365-2966.2011.18489.x}{\JournalTitle{\mnras},
  414, 1887}

\bibitem[{{Burstein} {et~al.}(1988){Burstein}, {Bertola}, {Buson}, {Faber}, \&
  {Lauer}}]{1988ApJ...328..440B}
{Burstein}, D., {Bertola}, F., {Buson}, L.~M., {Faber}, S.~M., \& {Lauer},
  T.~R. 1988, \href{http://dx.doi.org/10.1086/166304}{\JournalTitle{\apj}, 328,
  440}

\bibitem[{{Buzzoni} {et~al.}(2012){Buzzoni}, {Bertone}, {Carraro}, \&
  {Buson}}]{2012ApJ...749...35B}
{Buzzoni}, A., {Bertone}, E., {Carraro}, G., \& {Buson}, L. 2012,
  \href{http://dx.doi.org/10.1088/0004-637X/749/1/35}{\JournalTitle{\apj}, 749,
  35}

\bibitem[{{Cardelli} {et~al.}(1988){Cardelli}, {Clayton}, \&
  {Mathis}}]{1988ApJ...329L..33C}
{Cardelli}, J.~A., {Clayton}, G.~C., \& {Mathis}, J.~S. 1988,
  \href{http://dx.doi.org/10.1086/185171}{\JournalTitle{\apjl}, 329, L33}

\bibitem[{{Carraro} \& {Benvenuto}(2017)}]{2017ApJ...841L..10C}
{Carraro}, G., \& {Benvenuto}, O.~G. 2017,
  \href{http://dx.doi.org/10.3847/2041-8213/aa7131}{\JournalTitle{\apjl}, 841,
  L10}

\bibitem[{{Carter} {et~al.}(2011){Carter}, {Pass}, {Kennedy}, {Karick}, \&
  {Smith}}]{2011MNRAS.414.3410C}
{Carter}, D., {Pass}, S., {Kennedy}, J., {Karick}, A.~M., \& {Smith}, R.~J.
  2011,
  \href{http://dx.doi.org/10.1111/j.1365-2966.2011.18643.x}{\JournalTitle{\mnras},
  414, 3410}

\bibitem[{{Catelan}(2009)}]{2009ASSP....7..175C}
{Catelan}, M. 2009, \JournalTitle{Astrophysics and Space Science Proceedings},
  7, 175

\bibitem[{{Chantereau} {et~al.}(2015){Chantereau}, {Charbonnel}, \&
  {Decressin}}]{chantereau2015}
{Chantereau}, W., {Charbonnel}, C., \& {Decressin}, T. 2015,
  \href{http://dx.doi.org/10.1051/0004-6361/201525929}{\JournalTitle{\aap},
  578, A117}

\bibitem[{{Charbonnel}(2016)}]{2016EAS....80..177C}
{Charbonnel}, C. 2016, \href{http://dx.doi.org/10.1051/eas/1680006}{in EAS
  Publications Series, Vol.~80, EAS Publications Series, ed. E.~{Moraux},
  Y.~{Lebreton}, \& C.~{Charbonnel}}, 177

\bibitem[{{Chung} {et~al.}(2017){Chung}, {Yoon}, \& {Lee}}]{chung2017}
{Chung}, C., {Yoon}, S.-J., \& {Lee}, Y.-W. 2017,
  \href{http://dx.doi.org/10.3847/1538-4357/aa6f19}{\JournalTitle{\apj}, 842,
  91}

\bibitem[{{Code}(1969)}]{1969PASP...81..475C}
{Code}, A.~D. 1969,
  \href{http://dx.doi.org/10.1086/128809}{\JournalTitle{\pasp}, 81, 475}

\bibitem[{{Conroy} {et~al.}(2009){Conroy}, {Gunn}, \&
  {White}}]{2009ApJ...699..486C}
{Conroy}, C., {Gunn}, J.~E., \& {White}, M. 2009,
  \href{http://dx.doi.org/10.1088/0004-637X/699/1/486}{\JournalTitle{\apj},
  699, 486}

\bibitem[{{Dalessandro} {et~al.}(2012){Dalessandro}, {Schiavon}, {Rood},
  {Ferraro}, {Sohn}, {Lanzoni}, \& {O'Connell}}]{2012AJ....144..126D}
{Dalessandro}, E., {Schiavon}, R.~P., {Rood}, R.~T., {et~al.} 2012,
  \href{http://dx.doi.org/10.1088/0004-6256/144/5/126}{\JournalTitle{\aj}, 144,
  126}

\bibitem[{{D'Antona} {et~al.}(2016){D'Antona}, {Vesperini}, {D'Ercole},
  {Ventura}, {Milone}, {Marino}, \& {Tailo}}]{2016MNRAS.458.2122D}
{D'Antona}, F., {Vesperini}, E., {D'Ercole}, A., {et~al.} 2016,
  \href{http://dx.doi.org/10.1093/mnras/stw387}{\JournalTitle{\mnras}, 458,
  2122}

\bibitem[{{de Freitas Pacheco} \& {Barbuy}(1995)}]{1995A&A...302..718D}
{de Freitas Pacheco}, J.~A., \& {Barbuy}, B. 1995, \JournalTitle{\aap}, 302,
  718

\bibitem[{{De Propris}(2000)}]{2000MNRAS.316L...9D}
{De Propris}, R. 2000,
  \href{http://dx.doi.org/10.1046/j.1365-8711.2000.03696.x}{\JournalTitle{\mnras},
  316, L9}

\bibitem[{{Dorman} {et~al.}(1995){Dorman}, {O'Connell}, \&
  {Rood}}]{1995ApJ...442..105D}
{Dorman}, B., {O'Connell}, R.~W., \& {Rood}, R.~T. 1995,
  \href{http://dx.doi.org/10.1086/175428}{\JournalTitle{\apj}, 442, 105}

\bibitem[{{Dressler}(1980)}]{dressler1980}
{Dressler}, A. 1980,
  \href{http://dx.doi.org/10.1086/190663}{\JournalTitle{\apjs}, 42, 565}

\bibitem[{{Drinkwater} {et~al.}(2000){Drinkwater}, {Phillipps}, {Jones},
  {Gregg}, {Deady}, {Davies}, {Parker}, {Sadler}, \&
  {Smith}}]{2000A&A...355..900D}
{Drinkwater}, M.~J., {Phillipps}, S., {Jones}, J.~B., {et~al.} 2000,
  \JournalTitle{\aap}, 355, 900

\bibitem[{{Eisenhardt} {et~al.}(2007){Eisenhardt}, {De Propris}, {Gonzalez},
  {Stanford}, {Wang}, \& {Dickinson}}]{eisenhardt2007}
{Eisenhardt}, P.~R., {De Propris}, R., {Gonzalez}, A.~H., {et~al.} 2007,
  \href{http://dx.doi.org/10.1086/511688}{\JournalTitle{\apjs}, 169, 225}

\bibitem[{{Ferguson} \& {Sandage}(1988)}]{1988AJ.....96.1520F}
{Ferguson}, H.~C., \& {Sandage}, A. 1988,
  \href{http://dx.doi.org/10.1086/114903}{\JournalTitle{\aj}, 96, 1520}

\bibitem[{{Gil de Paz} {et~al.}(2007){Gil de Paz}, {Boissier}, {Madore},
  {Seibert}, {Joe}, {Boselli}, {Wyder}, {Thilker}, {Bianchi}, {Rey}, {Rich},
  {Barlow}, {Conrow}, {Forster}, {Friedman}, {Martin}, {Morrissey}, {Neff},
  {Schiminovich}, {Small}, {Donas}, {Heckman}, {Lee}, {Milliard}, {Szalay}, \&
  {Yi}}]{gil2007}
{Gil de Paz}, A., {Boissier}, S., {Madore}, B.~F., {et~al.} 2007,
  \href{http://dx.doi.org/10.1086/516636}{\JournalTitle{\apjs}, 173, 185}

\bibitem[{{Glazebrook} {et~al.}(2017){Glazebrook}, {Schreiber}, {Labb{\'e}},
  {Nanayakkara}, {Kacprzak}, {Oesch}, {Papovich}, {Spitler}, {Straatman},
  {Tran}, \& {Yuan}}]{2017Natur.544...71G}
{Glazebrook}, K., {Schreiber}, C., {Labb{\'e}}, I., {et~al.} 2017,
  \href{http://dx.doi.org/10.1038/nature21680}{\JournalTitle{\nat}, 544, 71}

\bibitem[{{Gratton} {et~al.}(2012){Gratton}, {Carretta}, \&
  {Bragaglia}}]{2012A&ARv..20...50G}
{Gratton}, R.~G., {Carretta}, E., \& {Bragaglia}, A. 2012,
  \href{http://dx.doi.org/10.1007/s00159-012-0050-3}{\JournalTitle{\aapr}, 20,
  50}

\bibitem[{{Greggio} \& {Renzini}(1990)}]{1990ApJ...364...35G}
{Greggio}, L., \& {Renzini}, A. 1990,
  \href{http://dx.doi.org/10.1086/169384}{\JournalTitle{\apj}, 364, 35}

\bibitem[{{Hammer} {et~al.}(2010{\natexlab{a}}){Hammer}, {Hornschemeier},
  {Mobasher}, {Miller}, {Smith}, {Arnouts}, {Milliard}, \&
  {Jenkins}}]{hammer2010}
{Hammer}, D., {Hornschemeier}, A.~E., {Mobasher}, B., {et~al.}
  2010{\natexlab{a}},
  \href{http://dx.doi.org/10.1088/0067-0049/190/1/43}{\JournalTitle{\apjs},
  190, 43}

\bibitem[{{Hammer} {et~al.}(2010{\natexlab{b}}){Hammer}, {Verdoes Kleijn},
  {Hoyos}, {den Brok}, {Balcells}, {Ferguson}, {Goudfrooij}, {Carter},
  {Guzm{\'a}n}, {Peletier}, {Smith}, {Graham}, {Trentham}, {Peng}, {Puzia},
  {Lucey}, {Jogee}, {Aguerri}, {Batcheldor}, {Bridges}, {Chiboucas}, {Davies},
  {del Burgo}, {Erwin}, {Hornschemeier}, {Hudson}, {Huxor}, {Jenkins},
  {Karick}, {Khosroshahi}, {Kourkchi}, {Komiyama}, {Lotz}, {Marzke},
  {Marinova}, {Matkovic}, {Merritt}, {Miller}, {Miller}, {Mobasher},
  {Mouhcine}, {Okamura}, {Percival}, {Phillipps}, {Poggianti}, {Price},
  {Sharples}, {Tully}, \& {Valentijn}}]{2010ApJS..191..143H}
{Hammer}, D., {Verdoes Kleijn}, G., {Hoyos}, C., {et~al.} 2010{\natexlab{b}},
  \href{http://dx.doi.org/10.1088/0067-0049/191/1/143}{\JournalTitle{\apjs},
  191, 143}

\bibitem[{{Han} {et~al.}(2007{\natexlab{a}}){Han}, {Podsiadlowski}, \&
  {Lynas-Gray}}]{han2007}
{Han}, Z., {Podsiadlowski}, P., \& {Lynas-Gray}, A.~E. 2007{\natexlab{a}},
  \href{http://dx.doi.org/10.1111/j.1365-2966.2007.12151.x}{\JournalTitle{\mnras},
  380, 1098}

\bibitem[{{Han} {et~al.}(2007{\natexlab{b}}){Han}, {Podsiadlowski}, \&
  {Lynas-Gray}}]{2007MNRAS.380.1098H}
---. 2007{\natexlab{b}},
  \href{http://dx.doi.org/10.1111/j.1365-2966.2007.12151.x}{\JournalTitle{\mnras},
  380, 1098}

\bibitem[{{Jester} {et~al.}(2005){Jester}, {Schneider}, {Richards}, {Green},
  {Schmidt}, {Hall}, {Strauss}, {Vanden Berk}, {Stoughton}, {Gunn},
  {Brinkmann}, {Kent}, {Smith}, {Tucker}, \& {Yanny}}]{2005AJ....130..873J}
{Jester}, S., {Schneider}, D.~P., {Richards}, G.~T., {et~al.} 2005,
  \href{http://dx.doi.org/10.1086/432466}{\JournalTitle{\aj}, 130, 873}

\bibitem[{{J{\o}rgensen} {et~al.}(2017){J{\o}rgensen}, {Chiboucas}, {Berkson},
  {Smith}, {Takamiya}, \& {Villaume}}]{jorgensen2017}
{J{\o}rgensen}, I., {Chiboucas}, K., {Berkson}, E., {et~al.} 2017,
  \href{http://dx.doi.org/10.3847/1538-3881/aa96a3}{\JournalTitle{\aj}, 154,
  251}

\bibitem[{{Karick} {et~al.}(2003){Karick}, {Drinkwater}, \&
  {Gregg}}]{2003MNRAS.344..188K}
{Karick}, A.~M., {Drinkwater}, M.~J., \& {Gregg}, M.~D. 2003,
  \href{http://dx.doi.org/10.1046/j.1365-8711.2003.06813.x}{\JournalTitle{\mnras},
  344, 188}

\bibitem[{{Le Cras} {et~al.}(2016){Le Cras}, {Maraston}, {Thomas}, \&
  {York}}]{2016MNRAS.461..766L}
{Le Cras}, C., {Maraston}, C., {Thomas}, D., \& {York}, D.~G. 2016,
  \href{http://dx.doi.org/10.1093/mnras/stw1024}{\JournalTitle{\mnras}, 461,
  766}

\bibitem[{{Lee} {et~al.}(2005{\natexlab{a}}){Lee}, {Joo}, {Han}, {Chung},
  {Ree}, {Sohn}, {Kim}, {Yoon}, {Yi}, \& {Demarque}}]{lee2005he}
{Lee}, Y.-W., {Joo}, S.-J., {Han}, S.-I., {et~al.} 2005{\natexlab{a}},
  \href{http://dx.doi.org/10.1086/428944}{\JournalTitle{\apjl}, 621, L57}

\bibitem[{{Lee} {et~al.}(2005{\natexlab{b}}){Lee}, {Ree}, {Rich}, {Deharveng},
  {Sohn}, {Rey}, {Yi}, {Yoon}, {Bianchi}, {Lee}, {Seibert}, {Barlow}, {Byun},
  {Donas}, {Forster}, {Friedman}, {Heckman}, {Jee}, {Jelinsky}, {Kim},
  {Madore}, {Malina}, {Martin}, {Milliard}, {Morrissey}, {Neff}, {Rhee},
  {Schiminovich}, {Siegmund}, {Small}, {Szalay}, {Welsh}, \& {Wyder}}]{Lee2005}
{Lee}, Y.-W., {Ree}, C.~H., {Rich}, R.~M., {et~al.} 2005{\natexlab{b}},
  \href{http://dx.doi.org/10.1086/422503}{\JournalTitle{\apjl}, 619, L103}

\bibitem[{{Linden} {et~al.}(2017){Linden}, {Pryal}, {Hayes}, {Troup},
  {Majewski}, {Andrews}, {Beers}, {Carrera}, {Cunha}, {Fern{\'a}ndez-Trincado},
  {Frinchaboy}, {Geisler}, {Lane}, {Nitschelm}, {Pan}, {Allende Prieto},
  {Roman-Lopes}, {Smith}, {Sobeck}, {Tang}, {Villanova}, \&
  {Zasowski}}]{Linden2017}
{Linden}, S.~T., {Pryal}, M., {Hayes}, C.~R., {et~al.} 2017,
  \href{http://dx.doi.org/10.3847/1538-4357/aa6f17}{\JournalTitle{\apj}, 842,
  49}

\bibitem[{{Markwardt} {et~al.}(2005){Markwardt}, {Tueller}, {Skinner},
  {Gehrels}, {Barthelmy}, \& {Mushotzky}}]{markwardt2005}
{Markwardt}, C.~B., {Tueller}, J., {Skinner}, G.~K., {et~al.} 2005,
  \href{http://dx.doi.org/10.1086/498569}{\JournalTitle{\apjl}, 633, L77}

\bibitem[{{Martin} {et~al.}(2005){Martin}, {Fanson}, {Schiminovich},
  {Morrissey}, {Friedman}, {Barlow}, {Conrow}, {Grange}, {Jelinsky},
  {Milliard}, {Siegmund}, {Bianchi}, {Byun}, {Donas}, {Forster}, {Heckman},
  {Lee}, {Madore}, {Malina}, {Neff}, {Rich}, {Small}, {Surber}, {Szalay},
  {Welsh}, \& {Wyder}}]{2005ApJ...619L...1M}
{Martin}, D.~C., {Fanson}, J., {Schiminovich}, D., {et~al.} 2005,
  \href{http://dx.doi.org/10.1086/426387}{\JournalTitle{\apjl}, 619, L1}

\bibitem[{{Michard} \& {Andreon}(2008)}]{2008A&A...490..923M}
{Michard}, R., \& {Andreon}, S. 2008,
  \href{http://dx.doi.org/10.1051/0004-6361:200810283}{\JournalTitle{\aap},
  490, 923}

\bibitem[{{Miglio} {et~al.}(2012){Miglio}, {Brogaard}, {Stello}, {Chaplin},
  {D'Antona}, {Montalb{\'a}n}, {Basu}, {Bressan}, {Grundahl}, {Pinsonneault},
  {Serenelli}, {Elsworth}, {Hekker}, {Kallinger}, {Mosser}, {Ventura},
  {Bonanno}, {Noels}, {Silva Aguirre}, {Szabo}, {Li}, {McCauliff}, {Middour},
  \& {Kjeldsen}}]{miglio2012}
{Miglio}, A., {Brogaard}, K., {Stello}, D., {et~al.} 2012,
  \href{http://dx.doi.org/10.1111/j.1365-2966.2011.19859.x}{\JournalTitle{\mnras},
  419, 2077}

\bibitem[{{Milone} {et~al.}(2015){Milone}, {Marino}, {Piotto}, {Renzini},
  {Bedin}, {Anderson}, {Cassisi}, {D'Antona}, {Bellini}, {Jerjen},
  {Pietrinferni}, \& {Ventura}}]{2015ApJ...808...51M}
{Milone}, A.~P., {Marino}, A.~F., {Piotto}, G., {et~al.} 2015,
  \href{http://dx.doi.org/10.1088/0004-637X/808/1/51}{\JournalTitle{\apj}, 808,
  51}

\bibitem[{{Morrissey} {et~al.}(2007){Morrissey}, {Conrow}, {Barlow}, {Small},
  {Seibert}, {Wyder}, {Budav{\'a}ri}, {Arnouts}, {Friedman}, {Forster},
  {Martin}, {Neff}, {Schiminovich}, {Bianchi}, {Donas}, {Heckman}, {Lee},
  {Madore}, {Milliard}, {Rich}, {Szalay}, {Welsh}, \&
  {Yi}}]{2007ApJS..173..682M}
{Morrissey}, P., {Conrow}, T., {Barlow}, T.~A., {et~al.} 2007,
  \href{http://dx.doi.org/10.1086/520512}{\JournalTitle{\apjs}, 173, 682}

\bibitem[{{Nelan} {et~al.}(2005){Nelan}, {Smith}, {Hudson}, {Wegner}, {Lucey},
  {Moore}, {Quinney}, \& {Suntzeff}}]{nelan2005}
{Nelan}, J.~E., {Smith}, R.~J., {Hudson}, M.~J., {et~al.} 2005,
  \href{http://dx.doi.org/10.1086/431962}{\JournalTitle{\apj}, 632, 137}

\bibitem[{{Norris}(2004)}]{norris2004}
{Norris}, J.~E. 2004,
  \href{http://dx.doi.org/10.1086/423986}{\JournalTitle{\apjl}, 612, L25}

\bibitem[{{O'Connell}(1999)}]{1999ARA&A..37..603O}
{O'Connell}, R.~W. 1999,
  \href{http://dx.doi.org/10.1146/annurev.astro.37.1.603}{\JournalTitle{\araa},
  37, 603}

\bibitem[{{O'Connell} {et~al.}(2005){O'Connell}, {Rosario}, {Schiavon},
  {Conselice}, {Gallagher}, \& {Wyse}}]{oconnell2005}
{O'Connell}, R.~W., {Rosario}, D.~J., {Schiavon}, R.~P., {et~al.} 2005, in
  Bulletin of the American Astronomical Society, Vol.~37, American Astronomical
  Society Meeting Abstracts, 1450

\bibitem[{{Pastorello} {et~al.}(2014){Pastorello}, {Forbes}, {Foster},
  {Brodie}, {Usher}, {Romanowsky}, {Strader}, \& {Arnold}}]{Pastorello2014}
{Pastorello}, N., {Forbes}, D.~A., {Foster}, C., {et~al.} 2014,
  \href{http://dx.doi.org/10.1093/mnras/stu937}{\JournalTitle{\mnras}, 442,
  1003}

\bibitem[{{Peacock} {et~al.}(2017){Peacock}, {Zepf}, {Kundu}, \&
  {Chael}}]{2017MNRAS.464..713P}
{Peacock}, M.~B., {Zepf}, S.~E., {Kundu}, A., \& {Chael}, J. 2017,
  \href{http://dx.doi.org/10.1093/mnras/stw2382}{\JournalTitle{\mnras}, 464,
  713}

\bibitem[{{Peng} \& {Nagai}(2009)}]{Peng2009}
{Peng}, F., \& {Nagai}, D. 2009,
  \href{http://dx.doi.org/10.1088/0004-637X/705/1/L58}{\JournalTitle{\apjl},
  705, L58}

\bibitem[{{Percival} \& {Salaris}(2011)}]{2011MNRAS.412.2445P}
{Percival}, S.~M., \& {Salaris}, M. 2011,
  \href{http://dx.doi.org/10.1111/j.1365-2966.2010.18066.x}{\JournalTitle{\mnras},
  412, 2445}

\bibitem[{{Piotto} {et~al.}(2005){Piotto}, {Villanova}, {Bedin}, {Gratton},
  {Cassisi}, {Momany}, {Recio-Blanco}, {Lucatello}, {Anderson}, {King},
  {Pietrinferni}, \& {Carraro}}]{2005ApJ...621..777P}
{Piotto}, G., {Villanova}, S., {Bedin}, L.~R., {et~al.} 2005,
  \href{http://dx.doi.org/10.1086/427796}{\JournalTitle{\apj}, 621, 777}

\bibitem[{{Piotto} {et~al.}(2007){Piotto}, {Bedin}, {Anderson}, {King},
  {Cassisi}, {Milone}, {Villanova}, {Pietrinferni}, \&
  {Renzini}}]{2007ApJ...661L..53P}
{Piotto}, G., {Bedin}, L.~R., {Anderson}, J., {et~al.} 2007,
  \href{http://dx.doi.org/10.1086/518503}{\JournalTitle{\apjl}, 661, L53}

\bibitem[{{Poole} {et~al.}(2008){Poole}, {Breeveld}, {Page}, {Landsman},
  {Holland}, {Roming}, {Kuin}, {Brown}, {Gronwall}, {Hunsberger}, {Koch},
  {Mason}, {Schady}, {vanden Berk}, {Blustin}, {Boyd}, {Broos}, {Carter},
  {Chester}, {Cucchiara}, {Hancock}, {Huckle}, {Immler}, {Ivanushkina},
  {Kennedy}, {Marshall}, {Morgan}, {Pandey}, {de Pasquale}, {Smith}, \&
  {Still}}]{2008MNRAS.383..627P}
{Poole}, T.~S., {Breeveld}, A.~A., {Page}, M.~J., {et~al.} 2008,
  \href{http://dx.doi.org/10.1111/j.1365-2966.2007.12563.x}{\JournalTitle{\mnras},
  383, 627}

\bibitem[{{Price} {et~al.}(2011){Price}, {Phillipps}, {Huxor}, {Smith}, \&
  {Lucey}}]{price2011}
{Price}, J., {Phillipps}, S., {Huxor}, A., {Smith}, R.~J., \& {Lucey}, J.~R.
  2011,
  \href{http://dx.doi.org/10.1111/j.1365-2966.2010.17862.x}{\JournalTitle{\mnras},
  411, 2558}

\bibitem[{{Recio-Blanco} {et~al.}(2006){Recio-Blanco}, {Aparicio}, {Piotto},
  {de Angeli}, \& {Djorgovski}}]{recio2006}
{Recio-Blanco}, A., {Aparicio}, A., {Piotto}, G., {de Angeli}, F., \&
  {Djorgovski}, S.~G. 2006,
  \href{http://dx.doi.org/10.1051/0004-6361:20053006}{\JournalTitle{\aap}, 452,
  875}

\bibitem[{{Roming} {et~al.}(2005){Roming}, {Kennedy}, {Mason}, {Nousek}, {Ahr},
  {Bingham}, {Broos}, {Carter}, {Hancock}, {Huckle}, {Hunsberger}, {Kawakami},
  {Killough}, {Koch}, {McLelland}, {Smith}, {Smith}, {Soto}, {Boyd},
  {Breeveld}, {Holland}, {Ivanushkina}, {Pryzby}, {Still}, \&
  {Stock}}]{2005SSRv..120...95R}
{Roming}, P.~W.~A., {Kennedy}, T.~E., {Mason}, K.~O., {et~al.} 2005,
  \href{http://dx.doi.org/10.1007/s11214-005-5095-4}{\JournalTitle{\ssr}, 120,
  95}

\bibitem[{{Salaris} {et~al.}(2016){Salaris}, {Cassisi}, \&
  {Pietrinferni}}]{salaris2016}
{Salaris}, M., {Cassisi}, S., \& {Pietrinferni}, A. 2016,
  \href{http://dx.doi.org/10.1051/0004-6361/201628181}{\JournalTitle{\aap},
  590, A64}

\bibitem[{{Schlafly} \& {Finkbeiner}(2011)}]{2011ApJ...737..103S}
{Schlafly}, E.~F., \& {Finkbeiner}, D.~P. 2011,
  \href{http://dx.doi.org/10.1088/0004-637X/737/2/103}{\JournalTitle{\apj},
  737, 103}

\bibitem[{{Schombert}(2016)}]{Schombert2016}
{Schombert}, J.~M. 2016,
  \href{http://dx.doi.org/10.3847/0004-6256/152/6/214}{\JournalTitle{\aj}, 152,
  214}

\bibitem[{{Smith} {et~al.}(2012){Smith}, {Lucey}, \&
  {Carter}}]{2012MNRAS.421.2982S}
{Smith}, R.~J., {Lucey}, J.~R., \& {Carter}, D. 2012,
  \href{http://dx.doi.org/10.1111/j.1365-2966.2012.20524.x}{\JournalTitle{\mnras},
  421, 2982}

\bibitem[{{Sweigart} {et~al.}(2002){Sweigart}, {Brown}, {Lanz}, {Landsman}, \&
  {Hubeny}}]{2002ASPC..265..261S}
{Sweigart}, A.~V., {Brown}, T.~M., {Lanz}, T., {Landsman}, W.~B., \& {Hubeny},
  I. 2002, in Astronomical Society of the Pacific Conference Series, Vol. 265,
  Omega Centauri, A Unique Window into Astrophysics, ed. F.~{van Leeuwen},
  J.~D. {Hughes}, \& G.~{Piotto}, 261

\bibitem[{{Tailo} {et~al.}(2017){Tailo}, {D'Antona}, {Milone}, {Bellini},
  {Ventura}, {Di Criscienzo}, {Cassisi}, {Piotto}, {Salaris}, {Brown},
  {Vesperini}, {Bedin}, {Marino}, {Nardiello}, \&
  {Anderson}}]{2017MNRAS.465.1046T}
{Tailo}, M., {D'Antona}, F., {Milone}, A.~P., {et~al.} 2017,
  \href{http://dx.doi.org/10.1093/mnras/stw2790}{\JournalTitle{\mnras}, 465,
  1046}

\bibitem[{{Tantalo} {et~al.}(1996){Tantalo}, {Chiosi}, {Bressan}, \&
  {Fagotto}}]{Tantalo1996}
{Tantalo}, R., {Chiosi}, C., {Bressan}, A., \& {Fagotto}, F. 1996,
  \JournalTitle{\aap}, 311, 361

\bibitem[{{Thomas} {et~al.}(2005){Thomas}, {Maraston}, {Bender}, \& {Mendes de
  Oliveira}}]{thomas2005}
{Thomas}, D., {Maraston}, C., {Bender}, R., \& {Mendes de Oliveira}, C. 2005,
  \href{http://dx.doi.org/10.1086/426932}{\JournalTitle{\apj}, 621, 673}

\bibitem[{{Thomas} {et~al.}(2010){Thomas}, {Maraston}, {Schawinski}, {Sarzi},
  \& {Silk}}]{thomas2010}
{Thomas}, D., {Maraston}, C., {Schawinski}, K., {Sarzi}, M., \& {Silk}, J.
  2010,
  \href{http://dx.doi.org/10.1111/j.1365-2966.2010.16427.x}{\JournalTitle{\mnras},
  404, 1775}

\bibitem[{{Yaron} {et~al.}(2008){Yaron}, {Kovetz}, \&
  {Prialnik}}]{2008IAUS..252..261Y}
{Yaron}, O., {Kovetz}, A., \& {Prialnik}, D. 2008,
  \href{http://dx.doi.org/10.1017/S1743921308022965}{in IAU Symposium, Vol.
  252, The Art of Modeling Stars in the 21st Century, ed. L.~{Deng} \& K.~L.
  {Chan}}, 261

\bibitem[{{Yi} {et~al.}(1997){Yi}, {Demarque}, \& {Oemler}}]{Yi1997}
{Yi}, S., {Demarque}, P., \& {Oemler}, Jr., A. 1997,
  \href{http://dx.doi.org/10.1086/304498}{\JournalTitle{\apj}, 486, 201}

\bibitem[{{Yi} {et~al.}(1999){Yi}, {Lee}, {Woo}, {Park}, {Demarque}, \&
  {Oemler}}]{Yi1999}
{Yi}, S., {Lee}, Y.-W., {Woo}, J.-H., {et~al.} 1999,
  \href{http://dx.doi.org/10.1086/306856}{\JournalTitle{\apj}, 513, 128}

\bibitem[{{Yi}(2008)}]{2008ASPC..392....3Y}
{Yi}, S.~K. 2008, in Astronomical Society of the Pacific Conference Series,
  Vol. 392, Hot Subdwarf Stars and Related Objects, ed. U.~{Heber}, C.~S.
  {Jeffery}, \& R.~{Napiwotzki}, 3

\bibitem[{{Yi} {et~al.}(2011){Yi}, {Lee}, {Sheen}, {Jeong}, {Suh}, \&
  {Oh}}]{2011ApJS..195...22Y}
{Yi}, S.~K., {Lee}, J., {Sheen}, Y.-K., {et~al.} 2011,
  \href{http://dx.doi.org/10.1088/0067-0049/195/2/22}{\JournalTitle{\apjs},
  195, 22}

\bibitem[{{Yi} {et~al.}(2005){Yi}, {Yoon}, {Kaviraj}, {Deharveng}, {Rich},
  {Salim}, {Boselli}, {Lee}, {Ree}, {Sohn}, {Rey}, {Lee}, {Rhee}, {Bianchi},
  {Byun}, {Donas}, {Friedman}, {Heckman}, {Jelinsky}, {Madore}, {Malina},
  {Martin}, {Milliard}, {Morrissey}, {Neff}, {Schiminovich}, {Siegmund},
  {Small}, {Szalay}, {Jee}, {Kim}, {Barlow}, {Forster}, {Welsh}, \&
  {Wyder}}]{2005ApJ...619L.111Y}
{Yi}, S.~K., {Yoon}, S.-J., {Kaviraj}, S., {et~al.} 2005,
  \href{http://dx.doi.org/10.1086/422811}{\JournalTitle{\apjl}, 619, L111}

\bibitem[{{York} {et~al.}(2000){York}, {Adelman}, {Anderson}, {Anderson},
  {Annis}, {Bahcall}, {Bakken}, {Barkhouser}, {Bastian}, {Berman}, {Boroski},
  {Bracker}, {Briegel}, {Briggs}, {Brinkmann}, {Brunner}, {Burles}, {Carey},
  {Carr}, {Castander}, {Chen}, {Colestock}, {Connolly}, {Crocker}, {Csabai},
  {Czarapata}, {Davis}, {Doi}, {Dombeck}, {Eisenstein}, {Ellman}, {Elms},
  {Evans}, {Fan}, {Federwitz}, {Fiscelli}, {Friedman}, {Frieman}, {Fukugita},
  {Gillespie}, {Gunn}, {Gurbani}, {de Haas}, {Haldeman}, {Harris}, {Hayes},
  {Heckman}, {Hennessy}, {Hindsley}, {Holm}, {Holmgren}, {Huang}, {Hull},
  {Husby}, {Ichikawa}, {Ichikawa}, {Ivezi{\'c}}, {Kent}, {Kim}, {Kinney},
  {Klaene}, {Kleinman}, {Kleinman}, {Knapp}, {Korienek}, {Kron}, {Kunszt},
  {Lamb}, {Lee}, {Leger}, {Limmongkol}, {Lindenmeyer}, {Long}, {Loomis},
  {Loveday}, {Lucinio}, {Lupton}, {MacKinnon}, {Mannery}, {Mantsch}, {Margon},
  {McGehee}, {McKay}, {Meiksin}, {Merelli}, {Monet}, {Munn}, {Narayanan},
  {Nash}, {Neilsen}, {Neswold}, {Newberg}, {Nichol}, {Nicinski}, {Nonino},
  {Okada}, {Okamura}, {Ostriker}, {Owen}, {Pauls}, {Peoples}, {Peterson},
  {Petravick}, {Pier}, {Pope}, {Pordes}, {Prosapio}, {Rechenmacher}, {Quinn},
  {Richards}, {Richmond}, {Rivetta}, {Rockosi}, {Ruthmansdorfer}, {Sandford},
  {Schlegel}, {Schneider}, {Sekiguchi}, {Sergey}, {Shimasaku}, {Siegmund},
  {Smee}, {Smith}, {Snedden}, {Stone}, {Stoughton}, {Strauss}, {Stubbs},
  {SubbaRao}, {Szalay}, {Szapudi}, {Szokoly}, {Thakar}, {Tremonti}, {Tucker},
  {Uomoto}, {Vanden Berk}, {Vogeley}, {Waddell}, {Wang}, {Watanabe},
  {Weinberg}, {Yanny}, {Yasuda}, \& {SDSS Collaboration}}]{2000AJ....120.1579Y}
{York}, D.~G., {Adelman}, J., {Anderson}, Jr., J.~E., {et~al.} 2000,
  \href{http://dx.doi.org/10.1086/301513}{\JournalTitle{\aj}, 120, 1579}

\end{thebibliography}

\end{document}